# Model Selection in Time Series Analysis: Using Information Criteria as an Alternative to Hypothesis Testing[1]

R. Scott Hacker

Jönköping International Business School, Jönköping University, Sweden

Scott.Hacker@jibs.hj.se

and

Abdulnasser Hatemi-J

Department of Economics, and Finance, UAE University

AHatemi@uaeu.ac.ae

**Abstract**

The issue of model selection in applied research is of vital importance. Since the true model in such research is not known, which model should be used from among various potential ones is an empirical question. There might exist several competitive models. A typical approach to dealing with this is classic hypothesis testing using an arbitrarily chosen significance level based on the underlying assumption that a true null hypothesis exists. In this paper we investigate how successful this approach is in determining the correct model for different data generating processes using time series data. An alternative approach based on more formal model selection techniques using an information criterion or cross-validation is suggested and evaluated in the time series environment via Monte Carlo experiments. This paper also explores the effectiveness of deciding what type of general relation exists between two variables (e.g. relation in levels or relation in first differences) using various strategies based on hypothesis testing and on information criteria with the presence or absence of unit roots.

[1] Different versions of this paper have been presented at the following universities: Queensland University of Technology, Australia; Deakin University, Australia; and UAE University, the UAE. The authors are indebted to participants for their comments. The usual disclaimer applies however.

## Model Selection in Time Series Analysis: Using Information Criteria as an Alternative to Hypothesis Testing

### 1. Introduction[2]

Modern time series analysis leads the researcher to consider a wide variety of data characteristics in determining whether a relation exists between variables. The researcher needs to be concerned about whether the variables are stationary or not, whether they each have a trend, and how many lags to include in examining the data. The researcher needs to also be concerned with whether a relation between the variables is apparent between the levels of the variables, perhaps through cointegration, or whether the relation is only apparent in first (or higher-order) differences. The existence of autocorrelation or heteroscedasticity and possibly correcting for these problems are also issues with which the researcher often needs to deal. All of this analysis with time series data renders the researcher to having to take into account a wide variety of potential models, and many of these models are not nested within others.

Researchers using time series data in economics and finance have usually proceeded in their analysis by estimating regression models and testing various hypotheses in a frequentist tradition, e.g. testing for a unit root, testing for cointegration, testing for autocorrelation, and so forth. Hypothesis testing has had a dual role in finance, economics and other scientific disciplines. First, it is used to give us a minimum degree of confidence about rejecting a null hypothesis on some parameter restriction(s) by controlling for type I error at a particular level which is arbitrarily chosen, such as 5%.[3] Second, it is used for model selection—if the null hypothesis is not rejected we often consider the model with the null hypothesis as being acceptable, whereas if the null hypothesis is rejected we find the model without the parameter restrictions in the null hypothesis as acceptable. The first role is formally the most appropriate use of hypothesis testing, but the model selection usage is, in our opinion, the usage that predominates in finance and economics.[4] This is clear since economists are prone to repeated testing of various models to arrive at an acceptable one that fits the data without patterns in the residuals and that is hopefully robust. That process of repeated testing and discarding of models based upon it clearly affects the statistical size of the associated tests.[5] With the final model that researchers end up with they often act as if they had that model in mind all along and go about their t-testing, F-testing, and confidence-interval creating as if there was no prior model selection process. With the prior model selecting however, the estimated degree of confidence that they have in the rejection of various hypotheses given by standard t-tests and F-tests can be very misleading.

---

[2] The authors would like to thank Professor Clive Granger for his useful comments on a previous version of this paper. Previous versions of this paper were also presented at Deakin University and Queensland University of Technology. We thank the participants at these seminars for their comments.

[3] Type II error is also a concern to the extent that within the traditional hypothesis testing approach practitioners are suggested to use the test with highest power for a given size of the test, but beyond that the level of power is often not of much concern.

[4] The use of hypothesis testing as a means for model selection has been criticized before. See for example Akaike (1974, 1981) and Sclove (1994).

The developments in computer science have made the ability to consider systematically a wider variety of models in searching for an optimal model more operational. Hypothesis testing has been used in this endeavor, through step-wise regression techniques for example. A more sophisticated usage of hypothesis testing to search through a multitude of models is provided by Hendry and Krolzig with their PC-GETS software (see Hendry (2000), Hendry and Krolzig (1999, 2001), Krolzig (2001), or Krolzig and Hendry (2001) for a description). Another means to systematically compare many models is through the minimization of an information criterion (recommended by Granger, King, and White, 1995, among others, with respect to time series data) or some other model selection criterion such as cross-validation estimates of prediction error. In our opinion the usage of information criteria has not taken place in empirical economic studies as extensively as it legitimately could be. Their primary usage seems to be in choosing lag lengths in time series models. There are several reasons behind this limited usage. First, as far as we know there have not been any systematic studies in the literature that can show whether using information criteria can improve inference as an alternative approach to hypothesis testing. Second, closely associated with the first reason, many are concerned that model selection through the use of an information criterion is simply a data mining tool, with questionable inference properties. This can lead to some deep philosophical discussion, but we think that some of the worst practices of data mining can actually be avoided by considering various alternative credible models and presenting the strength of evidence supporting each, which information criteria can provide. Third, although greater systematic usage of information criteria is accessible using current econometric software, it is not very convenient to the common practitioner using that software to make the necessary calculations for a large number of models.

There are a number of purposes of the current study. First, we show how information criteria and cross-validation may be used for model selection when using time series data. The study considers the investigation of a potential relation between two variables in each of those variables may or may not be stationary and may or may not have a trend. We limit the study to the situation in which at least one of the variables is determined independently from the other. Second, we compare the extensive usage of model selection in a time series environment to stylized mechanical uses of hypothesis testing to search for an acceptable time series model. Third, we show the performances of the information criterion, cross-validation, and hypothesis testing strategies in a variety of ways using response surfaces and the principal of minimax regret. We consider the ability of the various strategies (1) to choose the right model, (2) to choose the right relation type between the variables (in levels, only in first differences, or none, for example) regardless of whether exactly the right model is chosen, and (3) to have superior predictive properties. Predictive properties are focused on since the true model is hardly ever among those considered in economic studies given that our models are vast simplifications of complicated economic relations.[6] Fourth, we address the issue of how to consider the strength of evidence supporting one model versus another when using information criteria. This is an important issue since the typical use of information criteria is to select a model without consideration of how strongly an information criterion supports the model over others. We largely follow the procedures outlined in Burnham and Anderson (2002) in considering model uncertainty. In dealing with this issue we suggest the calculation of some weights that reflect the strength of evidence supporting the various models considered. These weights may also be used to average across parameter estimates of different models, resulting in estimates that are perhaps more precise. This can be especially important if an alternative model to the chosen model seems to be almost as equally supported by the data.

[5] This problem is known as the mass significance problem in the literature. It is most likely to be present if repeated testing includes nested models.

Our Monte Carlo simulations show that in the time series environment we are considering, minimization of an information criterion as a method for general model selection is preferable to hypothesis-testing strategies since the method often outperforms hypothesis testing in finding the correct model or relation type and in having superior predictive performance. Use of weights based on the information criterion values for the various considered models may be used to consider the strength of evidence supporting one model over another.

The rest of the paper is organized as follows. The next section presents the general class of time series models we are investigating and provides a taxonomy of various models within that class. In the third and fourth sections we present the information criteria, cross-validation, and stylized hypothesis-testing strategies we are considering. In the fifth section we outline our simulation design. In the sixth section we display response surfaces on the performance of the various strategies given some selected true data generating processes. In the seventh section we consider a wider variety of parameters for the true data generating processes and make comparisons using minimax regret concepts. In the eighth section we deal with the issue of strength of evidence in support of various models and provide a broader discussion of empirical use of information criteria in model selection. The final section provides the conclusion.

[6] Associated with this is the issue that the null hypothesis typically considered in hypothesis testing is probably not true *a priori*. For example testing the null hypothesis that the mean value of two groups is the same can be an unusual hypothesis to test, since the likelihood (in the Bayesian sense) that the two groups have exactly the same mean is very low.

**2. Taxonomies of generating processes**

The simulations in this paper focus on a single-equation relation between two variables, $Y$ and $Z$, in which $Z$ may affect $Y$, but $Y$ cannot affect $Z$. The most general equation that includes all the explanatory variables we are considering in determining $Y$ is

$$\Delta Y_{\mathrm{t}} = b_1 + b_2 t + b_3 Y_{t-1} + b_4 \Delta Y_{t-1} + b_5 \Delta Y_{t-2} + b_6 Z_t + b_7 \Delta Z_t + b_8 \Delta Z_{t-1} + b_9 \Delta Z_{t-2} + b_{10}(Y_{\mathrm{t-1}} - (c_1 + c_2 Z_{\mathrm{t-1}})) + u_t, \quad (1)$$

or equivalently

$$Y_{\mathrm{t}} = b_1 + b_2 t + (b_3 + 1) Y_{t-1} + b_4 \Delta Y_{t-1} + b_5 \Delta Y_{t-2} + b_6 Z_t + b_7 \Delta Z_t + b_8 \Delta Z_{t-1} + b_9 \Delta Z_{t-2} + b_{10}(Y_{\mathrm{t-1}} - (c_1 + c_2 Z_{\mathrm{t-1}})) + u_t, \quad (1')$$

where $t$ is the time subscript, $\Delta$ is the first-difference operator, $b_1, b_2, \ldots, b_{10}, c_1$ and $c_2$ are constants, and $u_t$ is an error term drawn from a standard normal distribution. We never include this general equation with all its coefficient parameters nonzero as a possible model in this paper, as such a model has no tradition of having theoretical interest, but we do consider many versions of this equation in which have zero constraints on various parameters. The variable $Z_t$ is generated according to processes based on the equation:

$$Z_t = m_1 + m_2 t + m_3 Z_{t-1} + \varepsilon_t, \quad (2)$$

where the $m_1$ and $m_2$ parameters take on zero or nonzero values, $m_3$ takes on values such that $0 < m_3 \leq 1$, and the error term $\varepsilon_t$ is drawn from a standard normal distribution, independent of $u_t$.

We will concentrate only on those cases in which the variables can have at maximum only one unit root. The choice of equation (1) as the most general equation for the $Y$ generating process is based on the following five issues we would like to consider in investigating how $Y$ and $Z$ are possibly related. First, the form of the potential level relation is relevant. If the two variables are stationary, a simple form of such a relation is

$$Y_t = b_1 + b_6 Z_t + u_t, \ b_6 \neq 0.^{7} \quad (3)$$

If the two variables are nonstationary, a level relation would need to be one of cointegration, and the simplest form of such a level relation consistent with Johansen (1988, 1991) testing is

$$\Delta Y_t = b_4\Delta Y_{t-1} + b_8\Delta Z_{t-1} + b_{10}(Y_{t-1} - (c_1 + c_2 Z_{t-1})) + u_t,\ b_{10} < 0,\ c_2 \neq 0. \quad (4)$$

Both equations (3) and (4) are included among the candidate models.

Second, we are interested in considering whether a relation between first differences in the variables exists, without there necessarily being a level relation. One form of such a relation would occur if equation (4) were the true data generating process with $b_8 \neq 0$ and $b_{10} = 0$. Since we are considering a relation in current levels in equation (3), it would seem odd not to consider the analogous situation with first differences. Therefore we also include

$$\Delta Y_t = b_7 \Delta Z_t + u_t,\ b_7 \neq 0 \quad (5)$$

among the candidate models. Third, we would like to allow for a possible time trend for *Y*, which leads to the consideration of a time variable, *t*, as an additional explanatory variable in equation (3), or an intercept term as an additional parameter to be estimated in equations (4) and (5).[8] Fourth, we would like to provide as alternatives to *Y* being related to *Z* some simple univariate processes for generating *Y*, with the four permutations of stationarity/nonstationarity and trend/no trend. The following processes provide the four permutations

$Y_t = Y_{t-1} + u_t$ (random walk: nonstationary, no trend) (6)

$Y_t = b_1 + Y_{t-1} + u_t,\ b_1 \neq 0$ (random walk with drift: nonstationary, with trend) (7)

$Y_t = b_1 + (b_3+1)Y_{t-1} + u_t,\ b_3+1 < 1$ (stationary, no trend) (8)

$Y_t = b_1 + b_2 t + (b_3+1)Y_{t-1} + u_t,\ b_3+1 < 1$ (stationary around trend).[9] (9)

---

[7] We could consider alternatively consider $Y_t$'s potential relation to $Z_{t-1}$, which would be equally simple, but we focus only on current levels relation in this paper to keep the discussion manageable.

[8] In equation (4) with cointegration, if *Z* has a trend, *Y* should have a trend also, even without an additional intercept term. However if $b_{10} = 0$ (no cointegration), inclusion of the additional intercept $b_1$ would be needed to induce a trend in *Y*.

[9] These are the four model alternative considered as plausible for simple unit root testing in Elder and Kennedy (2001).

Fifth, we would like to include various augmentation lags, by which we mean the additional lagged first difference variables found in the augmented Dickey and Fuller (1979) unit root test or the Johansen (1988, 1991) cointegration test. In the case of the unit root test this means inclusion of $\Delta Y_{t-1}, \Delta Y_{t-2},...$ as explanatory variables for $\Delta Y_t$, and in the case of the Johansen cointegration test this means the inclusion of $\Delta Y_{t-1}, \Delta Y_{t-2},...$ and $\Delta Z_{t-1}, \Delta Z_{t-2},...$ (with the same number of lags for ΔY and ΔZ), as explanatory variables for $\Delta Y_t$. Allowing for too many of these lags would make the simulation study too cumbersome, so we opted to have a maximum of two, which allows consideration of the effect of including augmentation lags but choosing the wrong lag length.

Comparing equation (1) to equations (3)–(9) shows that all of the variables in the latter equations show up in equation (1) and equation (1) does not include any extra explanatory variables except augmentation lags. Table 1 provides a taxonomy of the various models we wish to consider arising from various constraints on equation (1). For the model numbers given in that table, the digits before the decimal point distinguish a model, ignoring how many augmentation lags are included, and the digits after the decimal point indicate the number of augmentation lags included. For the model titles, the abbreviations AL(1) and AL(2) mean respectively inclusion of one augmentation lag and inclusion of two augmentation lags. In discussing models we sometimes refer to just the first digit, in which case we are referring to all models with that first digit, e.g. model 1 means model 1.00, model 1.01, and model 1.02 as a group. Note that in the way we refer to models, whether or not a parameter is zero is distinguishing, e.g. $Y_t = b_1 + u_t$ (model 5.00) is considered to be a different model from $Y_t = b_1 + b_6 Z_t + u_t$ (model 11.00) rather than a special case of model 11, since any shown parameters in Table 1 are assumed nonzero.

Models 1–4 provide the univariate situations noted in equations (6)–(9), with and without augmentation lags. Models 5 and 6 use the same equations as the univariate stationary models 3 and 4, but with the restriction that $b_3 = -1$, so the speed of convergence to equilibrium is immediate. We refer to this as "White Noise" since it produces the generating process $Y_t = b_1 + u_t$ when there is no time trend and no lag augmentations. When there is a deterministic trend term appended to this process we call the process "White noise around a trend" and when lag augmentations are appended we continue using these terms despite the fact that the lag augmentations formally make the process not white noise. The reason for inclusion of Models 5 and 6 is to decrease an otherwise high frequency of acceptance of an actual relationship between $Y$ and $Z$ in some situations when there is spurious correlation between them. More on this matter is discussed later in the paper.

**Table 1. Taxonomy of equations explaining $Y_t$**

| Model Number[a] | Relation type[b] | Title | Process for $Y_t$ generation[c] | Equivalent process for $\Delta Y_t$ generation[c] |
|---|---|---|---|---|
| 1.00 | D | Random Walk | $Y_t = Y_{t-1} + u_t$ | $\Delta Y_t = u_t$ (nothing to estimate) |
| 1.01 | D | Random Walk, no intercept, AL(1) | $Y_t = Y_{t-1} + b_4\Delta Y_{t-1} + u_t$ | $\Delta Y_t = b_4\Delta Y_{t-1} + u_t$ |
| 1.02 | D | Random Walk, no intercept, AL(2) | $Y_t = Y_{t-1} + b_4\Delta Y_{t-1} + b_5\Delta Y_{t-2} + u_t$ | $\Delta Y_t = b_4\Delta Y_{t-1} + b_5\Delta Y_{t-2} + u_t$ |
| 2.00 | D | Random Walk with drift | $Y_t = b_1 + Y_{t-1} + u_t$ | $\Delta Y_t = b_1 + u_t$ |
| 2.01 | D | Random Walk with drift, AL(1) | $Y_t = b_1 + Y_{t-1} + b_4\Delta Y_{t-1} + u_t$ | $\Delta Y_t = b_1 + b_4\Delta Y_{t-1} + u_t$ |
| 2.02 | D | Random Walk with drift, AL(2) | $Y_t = b_1 + Y_{t-1} + b_4\Delta Y_{t-1} + b_5\Delta Y_{t-2} + u_t$ | $\Delta Y_t = b_1 + b_4\Delta Y_{t-1} + b_5\Delta Y_{t-2} + u_t$ |
| 3.00 | D | Stationary around nonzero constant | $Y_t = b_1 + (b_3+1)Y_{t-1} + u_t$, $-1< b_3+1 < 1$, $b_3 + 1 \neq 0$ | $\Delta Y_t = b_1 + b_3Y_{t-1} + u_t$, $-2< b_3 < 0$, $b_3 \neq -1$ |
| 3.01 | D | Stationary around nonzero constant, univariate AL(1) | $Y_t = b_1 + (b_3+1)Y_{t-1} + b_4\Delta Y_{t-1} + u_t$, $-1< b_3+1 < 1$, $b_3 + 1 \neq 0$ | $\Delta Y_t = b_1 + b_3Y_{t-1} + b_4\Delta Y_{t-1} + u_t$, $-2< b_3 < 0$, $b_3 \neq -1$ |
| 3.02 | D | Stationary around nonzero constant, univariate AL(2) | $Y_t = b_1 + (b_3+1)Y_{t-1} + b_4\Delta Y_{t-1} + b_5\Delta Y_{t-2} + u_t$, $-1<b_3+1 < 1$, $b_3 + 1 \neq 0$ | $\Delta Y_t = b_1 + b_3Y_{t-1} + b_4\Delta Y_{t-1} + b_5\Delta Y_{t-2} + u_t$, $-2< b_3 < 0$, $b_3 \neq -1$ |
| 4.00 | D | Trend Stationary (without white noise) | $Y_t = b_1 + b_2t + (b_3+1)Y_{t-1} + u_t$, $-1< b_3+1 < 1$, $b_3 + 1 \neq 0$ | $\Delta Y_t = b_1 + b_2t + b_3Y_{t-1} + u_t$, $-2< b_3 < 0$, $b_3 \neq -1$ |
| 4.01 | D | Trend Stationary (without white nose), univariate AL(1) | $Y_t = b_1 + b_2t + (b_3+1)Y_{t-1} + b_4\Delta Y_{t-1} + u_t$, $-1< b_3+1 < 1$, $b_3 + 1 \neq 0$ | $\Delta Y_t = b_1 + b_2t + b_3Y_{t-1} + b_4\Delta Y_{t-1} + u_t$, $-2< b_3 < 0$, $b_3 \neq -1$ |
| 4.02 | D | Trend Stationary (without white noise), univariate AL(2) | $Y_t = b_1 + b_2t + (b_3+1)Y_{t-1} + b_4\Delta Y_{t-1} + b_5\Delta Y_{t-2} + u_t$, $-1< b_3+1 < 1$, $b_3 + 1 \neq 0$ | $\Delta Y_t = b_1 + b_2t + b_3Y_{t-1} + b_4\Delta Y_{t-1} + b_5\Delta Y_{t-2} + u_t$, $-2<b_3< 0$, $b_3 \neq -1$ |
| 5.00 | D | White noise | $Y_t = b_1 + u_t$ | $\Delta Y_t = b_1 - Y_{t-1} + u_t$, |
| 5.01 | D | White noise AL(1) | $Y_t = b_1 + b_4\Delta Y_{t-1} + u_t$, | $\Delta Y_t = b_1 - Y_{t-1} + b_4\Delta Y_{t-1} + u_t$, |
| 5.02 | D | White noise AL(2) | $Y_t = b_1 + b_4\Delta Y_{t-1} + b_5\Delta Y_{t-2} + u_t$, | $\Delta Y_t = b_1 - Y_{t-1} + b_4\Delta Y_{t-1} + b_5\Delta Y_{t-2} + u_t$, |
| 6.00 | D | White noise around trend | $Y_t = b_1 + b_2t +u_t$, | $\Delta Y_t = b_1 + b_2t - Y_{t-1} + u_t$ |
| 6.01 | D | White noise around trend AL(1) | $Y_t = b_1 + b_2t + b_4\Delta Y_{t-1} + u_t$, | $\Delta Y_t = b_1 + b_2t - Y_{t-1} + b_4\Delta Y_{t-1} + u_t$ |
| 6.02 | D | White noise around trend AL(2) | $Y_t = b_1 + b_2t + b_4\Delta Y_{t-1} + b_5\Delta Y_{t-2} + u_t$, | $\Delta Y_t = b_1 + b_2t - Y_{t-1} + b_4\Delta Y_{t-1} + b_5\Delta Y_{t-2} + u_t$, |
| 7.00 | B | Difference relation, no intercept | $Y_t = Y_{t-1} + b_7\Delta Z_t + u_t$ | $\Delta Y_t = b_7\Delta Z_t + u_t$ |
| 8.00 | B | Difference relation with intercept | $Y_t = b_1 + Y_{t-1} + b_7\Delta Z_t + u_t$ | $\Delta Y_t = b_1 + b_7\Delta Z_t + u_t$ |
| 9.01 | B | Difference Granger-causal model, no intercept, AL(1) | $Y_t = Y_{t-1} + b_4\Delta Y_{t-1} + b_8\Delta Z_{t-1} + u_t$ | $\Delta Y_t = b_4\Delta Y_{t-1} + b_8\Delta Z_{t-1} + u_t$ |
| 9.02 | B | Difference Granger-causal model, no intercept, AL(2) | $Y_t = Y_{t-1} + b_4\Delta Y_{t-1} + b_5\Delta Y_{t-2} + b_8\Delta Z_{t-1} + b_9\Delta Z_{t-2} + u_t$ | $\Delta Y_t = b_4\Delta Y_{t-1} + b_5\Delta Y_{t-2} + b_8\Delta Z_{t-1} + b_9\Delta Z_{t-2} + u_t$ |
| 10.01 | B | Difference Granger-causal model, with intercept, AL(1) | $Y_t = b_1 + Y_{t-1} + b_4\Delta Y_{t-1} + b_8\Delta Z_{t-1} + u_t$ | $\Delta Y_t = b_1 + b_4\Delta Y_{t-1} + b_8\Delta Z_{t-1} + u_t$ |
| 10.02 | B | Difference Granger-causal model, with intercept, AL(2) | $Y_t = b_1 + Y_{t-1} + b_4\Delta Y_{t-1} + b_5\Delta Y_{t-2} + b_8\Delta Z_{t-1} + b_9\Delta Z_{t-2} + u_t$ | $\Delta Y_t = b_1 + b_4\Delta Y_{t-1} + b_5\Delta Y_{t-2} + b_8\Delta Z_{t-1} + b_9\Delta Z_{t-2} + u_t$ |
| 11.00 | A | Current level relation | $Y_t = b_1 + b_6Z_t + u_t$ | $\Delta Y_t = b_1 - Y_{t-1} + b_6Z_t + u_t$ |
| 12.00 | A | Trend Current-level relation | $Y_t = b_1 + b_2t + b_6Z_t + u_t$ | $\Delta Y_t = b_1 + b_2t - Y_{t-1} + b_6Z_t + u_t$ |
| 13.01 | A | Error correction model, no intercept, AL(1) | $Y_t = Y_{t-1} + b_4\Delta Y_{t-1} + b_8\Delta Z_{t-1} + b_{10}(Y_{t-1} - (c_1 + c_2Z_{t-1})) + u_t$ | $\Delta Y_t = b_4\Delta Y_{t-1} + b_8\Delta Z_{t-1} + b_{10}(Y_{t-1} - (c_1 + c_2Z_{t-1})) + u_t$ |
| 13.02 | A | Error correction model, no intercept, AL(2) | $Y_t = Y_{t-1} + b_4\Delta Y_{t-1} + b_5\Delta Y_{t-2} + b_8\Delta Z_{t-1} + b_9\Delta Z_{t-2} +b_{10}(Y_{t-1} - (c_1 + c_2Z_{t-1})) + u_t$ | $\Delta Y_t = b_4\Delta Y_{t-1} + b_5\Delta Y_{t-2} + b_8\Delta Z_{t-1} + b_9\Delta Z_{t-2} +b_{10}(Y_{t-1} - (c_1 + c_2Z_{t-1})) + u_t$ |
| 14.01 | A | Error correction model, with intercept, AL(1) | $Y_t = b_1 + Y_{t-1} + b_4\Delta Y_{t-1} + b_8\Delta Z_{t-1} + b_{10}(Y_{t-1} - (c_1 + c_2Z_{t-1})) + u_t$ | $\Delta Y_t = b_1 + b_4\Delta Y_{t-1} + b_8\Delta Z_{t-1} + b_{10}(Y_{t-1} - (c_1 + c_2Z_{t-1})) + u_t$ |
| 14.02 | A | Error correction model, with intercept, AL(2) | $Y_t = b_1 + Y_{t-1} + b_4\Delta Y_{t-1} + b_5\Delta Y_{t-2} + b_8\Delta Z_{t-1} + b_9\Delta Z_{t-2} + b_{10}(Y_{t-1} - (c_1 + c_2Z_{t-1})) + u_t$ | $\Delta Y_t = b_1 + b_4\Delta Y_{t-1} + b_5\Delta Y_{t-2} + b_8\Delta Z_{t-1} + b_9\Delta Z_{t-2} +b_{10}(Y_{t-1} - (c_1 + c_2Z_{t-1})) + u_t$ |
| 15.01 | C | Stationary around lagged $\Delta Z$, no intercept, AL(1) | $Y_t = (b_3+1)Y_{t-1} + b_4\Delta Y_{t-1} + b_8\Delta Z_{t-1} + u_t$, $b_3+1 < 1$ | $\Delta Y_t = b_3Y_{t-1} + b_4\Delta Y_{t-1} + b_8\Delta Z_{t-1} + u_t$, $b_3<0$ |
| 15.02 | C | Stationary around lagged $\Delta Z$, intercept, AL(2) | $Y_t = (b_3+1)Y_{t-1} + b_4\Delta Y_{t-1} + b_5\Delta Y_{t-2} + b_8\Delta Z_{t-1} + b_9\Delta Z_{t-2} + u_t$, $b_3+1 < 1$ | $\Delta Y_t = b_3Y_{t-1} + b_4\Delta Y_{t-1} + b_5\Delta Y_{t-2} + b_8\Delta Z_{t-1} + b_9\Delta Z_{t-2} + u_t$, $b_3<0$ |
| 16.01 | C | Stationary around lagged $\Delta Z$, with intercept, AL(1) | $Y_t = b_1 + (b_3+1)Y_{t-1} + b_4\Delta Y_{t-1} + b_8\Delta Z_{t-1} + u_t$, $b_3+1 < 1$ | $\Delta Y_t = b_1 + b_3Y_{t-1} + b_4\Delta Y_{t-1} + b_8\Delta Z_{t-1} + u_t$, $b_3<0$ |
| 16.02 | C | Stationary around lagged $\Delta Z$, with intercept, AL(2) | $Y_t = b_1 + (b_3+1)Y_{t-1} + b_4\Delta Y_{t-1} + b_5\Delta Y_{t-2} + b_8\Delta Z_{t-1} + b_9\Delta Z_{t-2} + u_t$, $b_3+1 < 1$ | $\Delta Y_t = b_1 + b_3Y_{t-1} + b_4\Delta Y_{t-1} + b_5\Delta Y_{t-2} + b_8\Delta Z_{t-1} + b_9\Delta Z_{t-2} + u_t$, $b_3<0$ |

[a] In the text, references to a model number without the digits after the decimal point refers to the group of models with that number before the decimal point, e.g. model 1 refers to the group of modes 1.00, 1.01, and 1.02.
[b] The relation type between $Y$ and $Z$ is categorized as follows: A = relation in levels, B = relation only in first differences, C = mixed relation (relation neither purely in levels nor purely in differences, D = no relation among those considered).
[c] The $b_i$ ($i$ = 1,…,10) and $c_i$ ($i$ =1,2) parameters noted for each model are nonzero in each listed model.

Models 7 and 8 deal with the first-difference relation described in equation (5), without and with a nonzero intercept included (the nonzero intercept would induce a trending drift in the level of $Y$ unless counteracted exactly by an opposite trending drift in $Z$). Models 11 and 12 deal with the level relation in current values described in equation (3), without and with time included as an extra explanatory variable. Models 13 and 14 deal with the cointegrating relation described in equation (4), without and with an intercept included (a nonzero intercept could induce a trend as some other parameters approach zero), and with one or two augmentation lags. Models 9 and 10 are the same as models 13 and 14, except the speed of convergence term $b_{10}$ is set equal to zero so there is no cointegration. As such they become models of Granger causality in the first-differences of the variables. Models 15 and 16 also use restricted forms of the equations of models 13 and 14, with $c_1$ and $c_2$ equal to zero, so there is again no cointegration ($b_{10}$ in models 13 and 14 becomes $b_3$ in models 15 and 16).

We consider there to be a hierarchy in how variables are related to each other. If there is a level relation, i.e. a long-run relation between the variables, then some sort of relation in first differences is also implied. However, the reverse needs not hold—there may be only a short-run relation. Sometimes we have a relation between the variables that does not fit nicely within the category of a relation in levels or a relation in first differences, as in models 15 and 16. This leads us to categorize the models in Table 1 according to the following hierarchy on the relation type (as seen in the second column of the table):

A = $Y$ and $Z$ have a relation in levels.
B = $Y$ and $Z$ have a relation only in first differences.
C = $Y$ and $Z$ have a mixed relation, i.e. a relation that is neither purely in levels nor purely in first differences.
D = $Y$ and $Z$ have no relation (among those considered).

Relevant for some of the models in Table 1 is the corresponding data generating process for the explanatory variable $Z$. For example, cointegration between $Y$ and $Z$ requires a random walk (with or without drift) for both variables. The most general equation for the process generating $Z$ we consider is given by equation (2). Table 2 lists more specifically the types of processes we included in the simulations of this paper for generation of the $Z$ variable. These processes have the same form as models 1.00, 2.00, 3.00, and 4.00 for generating $Y$. We have chosen enough processes in generating $Z$ so that in connection with some of the models in Table 1, we can consider situations in which there is a relation between $Y$ and $Z$ where both are nonstationary with no trend, where both are nonstationary with a trend (through the drift), and where both are stationary, perhaps around a trend.

**Table 2. Taxonomy of models explaining $Z_t$**

| Process Name | Equation |
|---|---|
| Random Walk | $Z_t = Z_{t\text{-}1} + \varepsilon_t$ |
| Random Walk with drift | $Z_t = m_1 + Z_{t\text{-}1} + \varepsilon_t$, |
| Stationary around nonzero constant | $Z_t = m_1 + m_3 Z_{t\text{-}1} + \varepsilon_t$ , $m_3$<1 |
| Trend Stationary | $Z_t = m_1 + m_2 t + m_3 Z_{t\text{-}1} + \varepsilon_t$ , $m_3$<1 |

The $m_i$ ($i$ = 1, 2, 3) parameters noted are nonzero in each listed model.

## 3. Model selection using information criteria or cross-validation

In this section we present the information criteria we are considering in this paper for model selection, along with a cross-validation methodology. The information criteria we look at are the Akaike Information Criterion by Akaike (1973, 1974), the Sugiura (1978) and Hurvich and Tsai (1989) corrected Akaike Information Criterion, the McQuarrie, Shumway, and Tsai (1997) unbiased corrected Akaike Information Criterion, and the Schwarz Information Criterion by Schwarz (1978). Leave-one-out cross-validation is the investigated cross validation method. The Akaike information criterion is defined as

$$\mathrm{AIC} = T \ln\left(\frac{RSS}{T}\right) + 2(C+1)\,, \tag{10}$$

where RSS is the residual sum of squares for the estimated model, $T$ is the number of observations and $C$ is the number of estimated coefficient parameters including the intercept if it present. The corrected Akaike information criteria is

$$\mathrm{AICc} = T \ln\left(\frac{RSS}{T}\right) + \frac{2T(C+1)}{T-C-2}\,, \tag{11}$$

and the unbiased corrected Akaike information criterion using an unbiased variance estimate is defined as

$$\mathrm{AICu} = T \ln\left(\frac{RSS}{T-C}\right) + \frac{2T(C+1)}{T-C-2}\,. \tag{12}$$

The final criterion considered in this paper is the Schwarz information criterion, defined as

$$\mathrm{SIC} = T \ln\left(\frac{RSS}{T}\right) + (\ln T)C \tag{13}$$

which was introduced by Schwarz (1978) and an equivalent criterion was introduced by Akaike (1977, 1978).[10][11]

The goal of the Akaike information criterion and its adjusted versions, AICc and AICu, is optimal predictive performance in the domain of the original sample. AIC is asymptotically efficient (Shibata, 1980), providing a consistent estimator of predictive accuracy (Forster, 2001). AICc resolves some small-sample overfitting problems of AIC of AIC by using a direct estimate of the expected Kullback-Leibler distance in the derivation of the regression model, and otherwise it is asymptotically equivalent to AIC (McQuarrie and Tsai, 1998).[12] The use of the unbiased variance estimate in AICu is meant to deal with some overfitting properties of AIC and AICc asymptotically, but at the cost that it loses asymptotic equivalence with those earlier criteria (McQuarrie and Tsai, 1998).[13] Using a Bayesian derivation, the Schwarz information criterion instead aims to maximize the posterior probability of choosing the correct model if the correct model is among those choosable, and it is consistent in finding the true model under such circumstances. Rissanen (1978, 1983, 1989) provides another derivation of SIC based on minimizing the minimum description length.[14]

For the information criteria, the value of *C*, the number of unconstrained coefficient estimates used in the regression, is important, as a model is penalized the higher that number is. In the case of the error correction models (models 14 and 15) it is not so obvious whether that value should include the number of coefficient parameters estimated for the cointegrating vector, as we just use the residuals from the regression for the cointegrating vector as another explanatory variable. In this paper we consider the estimates for the intercept and slope coefficients in the cointegrating vector, $c_1$ and $c_2$, as among the estimated parameters to be included in the calculation of $C$. If we did not do that, then we would not take into account the extra fitting to the data that those parameter estimates provide.[15]

---

[10] The definitions for these information criteria in this paragraph differ in various sources. The definitions given here assume normally distributed errors with constant variance. Other definitions can differ by addition of a constant and/or multiplying by a constant, modifications which of course do not affect the results when finding a model that minimizes a particular criterion. In AIC, AICc, and AICu, the second term includes ($C$+1), representing the number of parameters to be estimated including the variance of the error term.

[11] The properties of different information criteria are investigated by Hacker and Hatemi-J (2008) via simulation methods.

[12] The Kullback-Leibler distance is discussed briefly in Section 8 of this paper.

[13] McQuarrie and Tsai (1998, pp. 32-33) note "that the probability that an efficient model selection criterion [e.g. AIC and AICc] will overfit by one particular extra variable is 0.1573, whereas consistent model selection criteria [e.g. SIC] overfit with probability 0…. The probability that AICu overfits by one particular extra variable is 0.0833, roughly halfway between 0 and 0.1573". The bracketed parts in the preceding quote are put in by the current authors.

[14] Another popular information criterion in time series analysis is the Hannan-Quinn (1979) one. Hatemi-J (2003, 2008) suggests another competitive criterion, which combines elements of the Hannan-Quinn information criterion and SIC. Performances on these measures are not included in this paper to save space.

The leave-one-out cross validation (CV) criterion is

$$CV = \frac{1}{T}\sum_{t=1}^{T}\left(\frac{Y_t - \hat{Y}_t}{1-h_t}\right)^2, \tag{14}$$

where $Y_t$ is the $t^{th}$ observation of $Y$, $\hat{Y}_t$ is the estimate for the $t^{th}$ value of $Y$ given the estimated model, $h_i$ is the $i^{th}$ diagonal element of the matrix $\chi(\chi'\chi)^{-1}\chi$, and $\chi$ is the data matrix for the independent variables in the model. The goal of CV is to estimate the average squared predictive error of an estimated equation without any assumptions about the true data generating process. The computation in (14) is equivalent to the mean of the square of the errors from the process in which for each of the observations the researcher first estimates an equation by ordinary least squares with the other observations, and using the resulting estimated equation measures the error in predicting the left-out observation (Wang, 2004).

For the purpose of model selection, the strategy when using AIC, AICc, AICu, SIC, or CV is to choose from a set of models that model which minimizes the given measure. To maintain comparability, all estimates of the $b_i$ parameters in each model of Table 1 use the same number of observations, even if more are available when less lags are needed. In our simulations using the information criteria or cross-validation, the estimated parameters in the cointegrating vector are the estimated parameters in model 11.00.

[15] Some may question whether the estimate for $c_1$ should be included in the sum of coefficient parameter estimates when $b_1$ is already included, since one may consider $b_1 - b_{10}c_1$ as being a single estimated intercept term to be included in the count. We find that the estimates for $b_1$ and $c_1$ provide different information, since the value for $m_1$ in the process generating $Z_t$ affects them differently, thereby warranting counting them separately. Note that the data generating processes in models 13 and 14 (15 and 16) are actually of the same form since both have an intercept: $-b_{10}c_1$ in the case of model 13 and $b_1 - b_{10}c_1$ in the case of model 14 (16). Making distinctions of these as being different models is in a sense artificial, based only on the fact that a two-stage process is used in which the cointegrating vector is estimated first, providing estimates for $c_1$ and $c_2$, and then estimating the error-correction model using the residuals from the first stage to get the other parameter estimates.

## 4. Strategies for choosing among dynamic specifications using hypothesis testing

There are many strategies suggested for determining whether a relation exists between two particular variables of interest with time series data. We have attempted to simulate what we think are the most common ways of approaching the issue using hypothesis testing, although in practice different econometricians will have their own favorite modifications to these strategies that they think represent the ideal way to tackle the problem. In this section we discuss the hypothesis-testing strategies used in our simulations.

First determine the univariate series status for each of the variables *Y* and *Z*, through the methodology displayed in Figure 2. Each variable can be determined to be random walk, random walk with drift, stationary around a constant, or stationary around trend with 0, 1 or 2 Dickey-Fuller Augmentations.

<u>If both are determined to be stationary (perhaps around trend)</u> ↓

Estimate $Y_t = b_1 + b_2 t + b_6 Z_t + u_t$, excluding $b_2 t$ if neither stationary around trend.

↓

Is autocorrelation indicated?[a]

No ↓ — Yes → Is the autocorrelation strong?[b]

- No → Re-estimate equation above using FGLS[e] ↲
- Yes → ↓ (to Estimate $\Delta Y_t = b_1 + b_7 \Delta Z_t + u_t$ below)

Can $b_6 = 0$ be rejected using *t* test?

↓ Yes — No → → → → ↓ (to Estimate $\Delta Y_t = b_1 + b_7 \Delta Z_t + u_t$ below)

Conclude current-level relation without trend (model 11) or with trend (model 12) depending on whether $b_2 t$ included.

<u>If both are determined to be random walk, both with or without drift</u> ↓

Test for cointegration between *Y* and *Z*: Can no cointegration be rejected?[c]

No ↓ (to Estimate $\Delta Y_t = b_1 + b_7 \Delta Z_t + u_t$ below) — ↓Yes

Conclude Error-Correction Relation

$$\Delta Y_t = b_1 + \sum_{k=1}^{K^*} \left( b_{3+k} \Delta Y_{t-k} + b_{7+k} \Delta Z_{t-k} \right) + b_{10} \varepsilon_{t-1} + u_t$$

where $\varepsilon_t = Y_t - c_1 - c_2 Z_t$, (model 13 or 14, depending on $b_1$ constrained or not to 0, with the augmentation lag = $K^*$)[d]

<u>If neither of the two statements to left is true</u> ↓ (to Estimate $\Delta Y_t = b_1 + b_7 \Delta Z_t + u_t$ below)

Estimate $\Delta Y_t = b_1 + b_7 \Delta Z_t + u_t$, excluding $b_1$ if neither variable displays trend.

↓

Is no autocorrelation indicated[a] and can $b_7 = 0$ be rejected by a standard *t* test?

↓ Yes — No↓

Yes: Conclude this is a model of a current 1st-difference relation (model 7 or 8 depending on whether $b_1$ constrained to 0)

No: Estimate $\Delta Y_t = b_1 + \sum_{k=1}^{K} \left( b_{3+k} \Delta Y_{t-k} + b_{7+k} \Delta Z_{t-k} \right) + u_t$, excluding $b_1$ if no trend or drift in both *Y* and *Z* and using SIC to determine *K*.

↓

Can $b_8 = b_9 = 0$ be rejected?[f]

↓No — ↓Yes

No: Conclude no relation between the variables, with $Y_t$ simply following its univariate series process determined in Figure 2 (model 1, 2, 3, 4, 5, or 6, with *Y*'s associated augmentation lag from Figure 2).

Yes: Conclude difference Granger causal relation (model 9 or 10 depending on $b_1$ constrained or not to 0, with the augmentation lag = *K*)

**Figure 1. Strategies based on hypothesis testing**

Notes: [a]Using Breusch-Godfrey test. [b]Autocorrelation is considered not strong for these purposes if the Durbin-Watson statistic ≥ $R^2$, as suggested by Maddala (1988) and if the coefficient estimate on the lagged residual in a Breusch-Godfrey test for autocorrelation is less than one. [c]Using either the Engle-Granger technique (in strategy EG) or the Johansen technique (in strategy Jo). [d]Constraint of $b_1 = 0$ if no drifts in random walks of *Y* and Z; determination of $K^*$ discussed in text. [e]Feasible Generalized Least Squares, in this case utilizing 2-step Cochrane-Orcutt adjustment. [f]If $b_9$ not estimated, then the question is "Can $b_8 = 0$ be rejected?"

Our generalization of using hypothesis testing to determine whether a relation exists between two variables and the structure of that relation is summarized in Figure 1. It includes checking for the stationarity status of each variable, the results of which determine subsequently what types of relationships between the variables are investigated: a simple level relationship, a relationship in current first differences of the variables, a long-run level relationship through cointegration, or a Granger-causal relationship in first differences of the variables. Figure 1 also adjusts the models tested according to whether or not the two variables have a trend (known *a priori* or suggested by the data), and it includes some testing for autocorrelation which can affect the final model selected. This figure represents only a caricature of the model selection process using hypothesis testing and is of course not fully representative of applications in reality. Usually applied econometricians supplement the mechanical process with looking for patterns in the raw data and residuals when making decisions about the usefulness of various models. We focus on the mechanical aspect as it is that part of the model selection process which can be subjected to simulation.

Notable in this figure is that sometimes more tests are done than may occur in reality. Many practitioners for example may stop looking for a relation between variables after testing for cointegration and not finding it. However, such a conclusion does not lead to a clear accepted alternative model for the data generating process and further testing is needed to determine an acceptable model. Another potential aberration from practice in reality is that testing for Granger causality in the first differences in the variables (the last test in the lower right of Figure 1) may not be the natural thing to do for many practitioners after failing to find a relation between the current first-difference of the variables. We include it since a Granger-causal relation in first differences needs to be selectable by the information criteria as an alternative to cointegration, and if information criteria can select Granger causal relations in first differences, then we want hypothesis testing to have the same opportunity in the simulations.

At the top of Figure 1 there is for each variable a determination of the *univariate series status* (random walk, random walk with drift, stationary around a constant, or stationary around trend with 0, 1 or 2 Dickey-Fuller Augmentations). At the bottom left of Figure 1, one of a number of *univariate series processes* can be concluded (models 1-6 with the distinguishing 0, 1, or 2 augmentation lags, a classification which has slightly more categories than univariate series status if one ignores the distinguishing augmentation lags). How the univariate series status and the univariate series process are determined is displayed in Figure 2 (where variable $x$ refers to variable $Y$ or $Z$ in Figure 1, as appropriate). The strategy in Figure 2 starts with a unit root test using the appropriate Dickey-Fuller critical value,[16] including time as a explanatory variable if there is known to be a trend in the variable or if the variable's trend status is unknown, and then proceeds to test for a trend if the trend status is unknown. If stationarity around a constant or around a trend is determined, then further testing for white noise (*b* = -1) is performed, which is relevant for determining the univariate series process.[17]

**If it is known that no trend exists:**

Estimate $\Delta x_t = a + bx_{t-1} + \sum_{k=1}^{K} \lambda_k \Delta x_{t-k} + e_t$

↓

Can *b* = 0 be rejected using t-statistic and Dickey-Fuller critical value?→ Yes: Decide stationary around a constant[a]

↓No

Decide random walk (Model 1, if dealing with *Y*)

**If it is known that a trend does exist:**

Estimate $\Delta x_t = a + bx_{t-1} + ct + \sum_{k=1}^{K} \lambda_k \Delta x_{t-k} + e_t$.

↓

Can *b* = 0 be rejected using t-statistic and Dickey-Fuller critical value?→ Yes: Decide stationary around trend[b]

↓No

Decide random walk with drift (Model 2, if dealing with *Y*)

**If trend status is unknown:**

Estimate $\Delta x_t = a + bx_{t-1} + ct + \sum_{k=1}^{K} \lambda_k \Delta x_{t-k} + e_t$.

↓

Can *b* = 0 be rejected using t-statistic and Dickey-Fuller critical value?→ Yes

↓

Can *c* = 0 be rejected using *t* statistic and the usual *t* distribution?

↓ No: Decide stationary around a constant[a]

↓ Yes: Decide stationary around trend[b]

No ↓

Estimate $\Delta x_t = a + \sum_{k=1}^{K} \lambda_k \Delta x_{t-k} + e_t$ → Can *a* = 0 be rejected using *t* statistic and the usual *t* distribution?

↓ No: Decide random walk (Model 1 if dealing with *Y*)

↓ Yes: Decide random walk with drift. (Model 2 if dealing with *Y*)

**Figure 2. Determining univariate series status and the univariate process for a variable *x***

Notes: [a]In the case of determining the accepted univariate process for *Y*, the hypothesis *b* = -1 is also tested. If it can be rejected model 3 (stationary around nonzero constant) is accepted, otherwise model 5 (white noise) is accepted.
[b]In the case of determining the accepted univariate process for *Y*, the hypothesis *b* = -1 is tested. If it can be rejected model 4 (trend stationary) is accepted, otherwise model 6 (white noise around trend) is accepted.

[16] The Dickey-Fuller critical values discussed are not exactly those given in the original Dickey and Fuller (1979) article. They are instead the critical values based on the formula and estimates for the response surface from unit root testing given in Mackinnon (1991). The critical values derived from Mackinnon are more flexible in number of observations and are arguably more precise than the original Dickey and Fuller critical values.

[17] The strategies in Figure 2 (except for the additional test for white noise noted in the footnotes of that table) we attribute to Elder and Kennedy (2001). The additional test on whether or not *b*=-1 is likely not to occur much in practice, but we think it is a harmless extension of the hypothesis testing strategy, and we include it to provide more comparability to the model selection strategy we are suggesting using information criteria or cross-validation.

In both figures the used number of augmentation lags, $K$ and $K^*$, are determined through minimization of the final prediction error using an unbiased estimator for variance,[18] with a maximum of two lags (as discussed shortly, a multivariate version of SIC may be used instead for determining $K^*$ depending on the type of cointegration test used). These two figures indicate that all of the models in Table 1 except models 15 and 16 are choosable according to our stylized hypothesis testing strategy. In our simulations, models 15 and 16 are choosable using information criteria, although they are not allowed as true data generating processes. Since they are not allowed as true data generating processes, we consider this asymmetry in what the hypothesis tests can choose and what the information criteria can choose as rather harmless for comparison of the techniques, except perhaps in consideration of predictive performance in some cases.

Figures 1 and 2 otherwise should be largely self-explanatory, although some discussion of the details surrounding possible cointegration situations in Figure 1 is necessary. Such a discussion is provided below, and that is followed by further discussion about Figure 2's handling of unit root testing when trend status is unknown.

In Figure 1, a cointegration test is performed if $Y$ and $Z$ are both determined to have a random walk, both with or without a drift. One of two types of cointegration tests is used – the Engle and Granger (1987) test (in which case Figure 1 describes what we refer to as strategy EG) or the test suggest by Johansen (1988, 1991) and Johansen and Juselius (1990) (in which case Figure 1 describes what we refer to as strategy Jo). The Engle and Granger test involves testing whether $\gamma_2 = 0$ can be rejected in the following equation after it has been estimated:

$$\Delta\hat{\varepsilon}_t = \gamma_1 + \gamma_2\hat{\varepsilon}_{t-1} + v_t, \tag{15}$$

where $\hat{\varepsilon}_t$ represents the residual at time $t$ from a simple linear regression of $Y$ on $Z$ (including an intercept), $\gamma_1$ and $\gamma_2$ are constant parameters, and $v_t$ is an error term. The t-value associated with the $\gamma_2$ estimate is compared to a critical value from MacKinnon (1991) to perform this test. If cointegration is concluded using this test an estimated form of the vector error correction equation

[18] The final prediction error was proposed by Akaike (1969,1970). The final prediction error using an unbiased estimator for variance was presented on pages 33-34 in McQuarrie and Tsai (1998) and is given by the formula $FPE_U = [RSS/(T-C)]\cdot[(T+C+1)/(T-C-1)]$. If $K$ and $K^*$ were instead chosen using one of the information criteria covered in this study, different conclusions on comparisons of information criteria and the hypothesis-testing techniques could result.

$$\Delta Y_t = b_1 + \sum_{k=1}^{K^*}(b_{3+k}\Delta Y_{t-k} + b_{7+k}\Delta Z_{t-k}) + b_{10}\hat{\varepsilon}_{t-1} + u_t \tag{16}$$

is accepted (excluding $b_1$ if no drifts are involved in the random walks of the variables, and with $K^*$ based on minimization of SIC) and is used for considering the predictive capabilities of the strategy.

The cointegration test following Johansen (1988, 1991) and Johansen and Juselius (1990) starts by estimating the following vector error correction equation

$$\Delta X_t = \Phi + \alpha[\rho', \beta']\begin{bmatrix} 1 \\ X_{t-1} \end{bmatrix} + \sum_{k=1}^{K^*}\Gamma_k \Delta X_{t-k} + w_t, \tag{17}$$

where $\Delta X_t \equiv [\Delta Y_t, \Delta Z_t]'$ and where $\Phi$ is excluded if no drifts are involved in the random walks of the variables. $K^*$ is determined by minimization of a multivariate version of the Schwarz information criterion in this case, again with a maximum of two lags. The test for cointegration is based on the trace of the estimated $\alpha$, using asymptotic critical values from the program associated with Mackinnon, Haug, and Michelis (1999). If cointegration is concluded, the single-equation error correction equation subsequently concluded is that using the same $K^*$ used in the cointegration test and with $b_1 = 0$ if $\Phi$ was constrained to be a zero vector in the cointegration test. However, in evaluating the predictive ability of the accepted model the estimated vector error correction model is used.[19]

How to proceed when the trend status of the variable is unknown is a contentious issue. As shown in Figure 2, we have chosen a method in which under such circumstances the unit root is tested with a deterministic time trend included (despite the low test power this induces if it is unnecessary) since the actual size of the unit root test should then be close to its nominal size regardless of the true trend status of the variable, and then the trend is tested with the form of that test based on the result of the unit root test. This method, suggested in Elder and Kennedy (2001), is also easy to apply, using tests widely available in standard econometrics textbooks. However, some researchers may prefer to do sequential testing of the unit root, which includes testing for the unit root perhaps again after further testing suggests there is no trend in the variable.[20] Other researchers may prefer first to perform a test for the trend that is robust to the stationarity status of the variable examined and then perform a test for the unit root with inclusion a deterministic trend in that test based the results of the previous test for the trend.[21] In any case our simulations deal with situations both where the trend status is known and where it is unknown. The simulations of situations in which the trend status is known we find interesting in their own right and also have the additional benefit of providing information that avoids the contentious issue of how to proceed when trend status is unknown.

[19] In the Johansen methodology, reduced rank regressions are used. We calculate $E[\Delta X_t] = \pi r_{\Omega t} + \Lambda\Omega_t$, where $\Omega_t$ includes all the augmentation lag variables and constants showing up outside the cointegrating vector at time $t$, $\Lambda$ is the estimated parameters from a preliminary regression of $\Delta X$ on $\Omega$, $r_{\Omega t}$ is the vector of residuals from the preliminary regression of the variables of the cointegrating vector on $\Omega_t$, and $\pi$ is the estimate of $\alpha[\rho', \beta']$.

## 5. Simulation Design

In the next two sections we present results from Monte Carlo simulations to consider the performance of various strategies in choosing models from the ones listed in Table 1.[22] The results for every set of true parameter values we use are based on 10,000 Monte Carlo simulations using 50 observations, with 100 presample observations generated to reduce the effect of startup values on the results. The error term $u_t$ used in generating the $Y_t$ series and the error term $\varepsilon_t$ used in generating the $Z_t$ series are each independently drawn from the standard normal distribution. The starting lag values for $Y_t$ and $Z_t$ in the presample are zeros and the last presample observations are used for initial lags of the variables in the used 50 observations. In moving from one observation to the next, the time variable $t$ increases by a discrete unit of one. Formally the simulations converted the single equation processes denoted in Tables 1 and 2 into the associated vector autoregressive (VAR) process and generated the data based on the reduced form for the VAR system. The details of this conversion are given in the appendix.

Each of the models in Table 1 is estimated with ordinary least squares using the 50 observations. The number of lags in the explanatory variables does not affect the number of observations in these estimates, as the presample observations may be used for the lags. One may think of the presample observations used in this way as actually being part of the actual sample, but are used only for lags. We have two augmentation lags as the maximum considered in the simulations, so with the two-lag first-difference variables we require three extra lags (e.g. $\Delta Y_{t-2} = Y_{t-2} - Y_{t-3}$ requires information on $Y_{t-3}$ ). In models with an error-correction relation, how the model estimates are handled depends on which cointegration testing technique is used. If the Engle and Granger (1987) cointegration is being used, the parameters for the potential cointegrating vector are estimated first in a separate linear regression of $Y_t$ on $Z_t$ (including intercept),[23] and the lagged residuals $\hat{\varepsilon}_{t-1}$, are used instead of $Y_{t-1} - (c_1 + c_2 Z_{t-1})$ when estimating the other parameters in the error correction model. If instead the cointegrating testing methodology advocated by Johansen (1988, 1991) and Johansen and Juselius (1990) is being used, then the vector error correction model with cointegration restrictions is estimated using the reduced-rank regression technique associated with that testing and is used for predictive purposes instead of the corresponding single-equation error-correction models found in Table 1.

[20] Various versions of this process have been suggested in the past, for example in Dolado, Jenkinson, and Sosvilla-Rivero (1990), Enders (2004), and Ayat and Burridge (2000). Strategies of this type of course make rejecting the unit root more likely, but at the expense of over-rejection of the null hypothesis of a unit root for the given nominal size when the unit root exists. Hacker and Hatemi-J (2010) show the degree to which size is distorted using the Enders (2004) strategy and the gains from using the Elder and Kennedy (2001) strategy instead.

[21] A unit root test which is robust to trend status is provided in Vogelsang (1998) and Bunzel and Vogelsang (2005).

[22] The simulations are performed through a GAUSS program.

Given a sample of the generated data we wish to find the strategies that are most helpful in providing an estimated model close to the one that actually generated the data. Each strategy based upon finding the model that minimizes one of the information criteria is referred to by the acronym for that information criterion. The strategy finding the model that minimizes the leave-one-out cross-validation measure is referred to as CV. The strategy based upon hypothesis testing as outlined in section 4 using the Engle-Granger test for cointegration (if any testing for cointegration is performed) is referred to as strategy EG. The strategy based upon hypothesis testing as outlined in section 4 using the Johansen (1988, 1991) test for cointegration (if any testing for cointegration is performed) is referred to as strategy Jo. We append to EG or Jo the nominal significance level used in each test in the hypothesis testing strategy, e.g. if strategy EG is used with a 5% nominal significance level being used on every test, then we refer to it as EG-5%. We also consider some mixed nominal significance levels. The denotation EG-10/5 represents the EG strategy in which all significance tests are performed at the 5% significance level except the unit root test which is performed at the 10% level. Jo-10/5 represents the same strategy except the Johansen cointegration test is performed when a cointegration test is deemed necessary. The use of such a mixed nominal significance level is to deal partly with the issue that augmented Dickey-Fuller tests often have low power.[24]

---

[23] When estimating the potential cointegrating vector coefficient parameters, the three presample observations used for lags elsewhere in the estimated model are used, as they would likely be used in practice to provide a small improvement to the estimates and they do not diminish the comparability of the models in Figure 1.

[24] To get the size on various tests to match their nominal sizes when pretesting exists, some adjustments should be made. Maddala and Kim (1998) for example note that the pre-testing literature suggests that the nominal size that should be used with unit root tests should be as high as 25% so that sizes on later tests are more accurate. We do not allow for that much correction as we think more standard levels of 5% and 10% are more commonly used in practice despite the problems with matching nominal and actual sizes.

Each of the strategies simulated must choose from the set of models provided in Table 1. Sometimes we limit the models from Table 1 that may be chosen based upon pre-knowledge about whether or not there is a trend in one of the variables, *Y* and *Z*. If we wish to consider only models in which there is no trend in either variable, then we limit the choosable models to the odd-numbered models: 1, 3, 5, 7, 9, 11, 13, and 15. If we wish to consider only models in which there is a trend in one of the variables, then we limit the choosable models to the even-numbered models: 2, 4, 6, 8, 10, 12, 14, and 16.

## 6. Performance of various model selection strategies for some selected true data generating processes

Evaluating the performance of model selection strategies is a difficult task, as there is an infinite number of possible "true" data generating processes. In this section we focus on various selected types of data generating processes and allow one parameter to change in each to generate response surfaces. This allows us to present what are some of the important strengths and weaknesses in each of the model selection strategies. We also introduce in this section three different types of performances to be considered for the response surfaces. These types of performances are considered again over a multitude of "true" data generating processes in the next section.

The first type of performance we investigate is the ability of the various model selection strategies to choose the correct model when it is among the possible models to be chosen. For five model selection strategies (AIC, SIC, CV, and Jo-10%, Jo-5%, ) and six different types of data generating processes for *Y*, Figure 3 compares the response surfaces on the frequency of choosing the correct model (including correct number of augmentation lags) when there is known to be no trends in *Y* and *Z* so models with a trend are not choosable. In each part of Figure 3, the lowest value the varying parameter (e.g. $b_6$ in Figure 3a) takes on is 0.00001, and in Figure 3e the varying parameter, $-b_3$, is highest at 0.99999, so for the whole response surface the true data generating process is associated with only one model from Table 1. To illuminate how this figure should be understood, consider for example Figure 3a, in which the data is generated according to the equation $Y_t = 1 + b_6 Z_t + u_t$, $b_6 \geq 0.00001$with *Z* generated by the equation $Z_t = 1 + 0.5 Z_{t-1} + \varepsilon_t$. Formally for all points along the response surface, the model is model 11.00 from Table 1 ( $Y_t = b_1 + b_6 Z_t + u_t$ with $b_1 \neq 0$, $b_6 \neq 0$) and we can appropriately say that what is being measured on the figure along the whole response surface is the frequency of choosing model 11.00, the correct model.

What we see from the response surfaces in Figure 3a is that all the model selection strategies converge to choosing the correct model as $b_6$ increases, as we would expect, and that the model selection criteria SIC, AIC, and CV have roughly similar performance to that of the hypothesis-testing strategies at choosing the correct model in this situation. The performance of SIC seems to closely follow the performance of Jo-5% for low $b_6$ values and SIC performs best among all the shown strategies when $b_6 \geq 0.6$. AIC is the most successful at choosing the correct model if $0 < b_6 \leq 0.5$, but that also leads to the undesirable characteristic for AIC that if $b_6$ is zero, AIC would be the least successful in choosing the model $Y_t = b_1 + u_t,\ b_1 \neq 0$.

3a: $Y_t = 1 + b_6 Z_t + u_t$

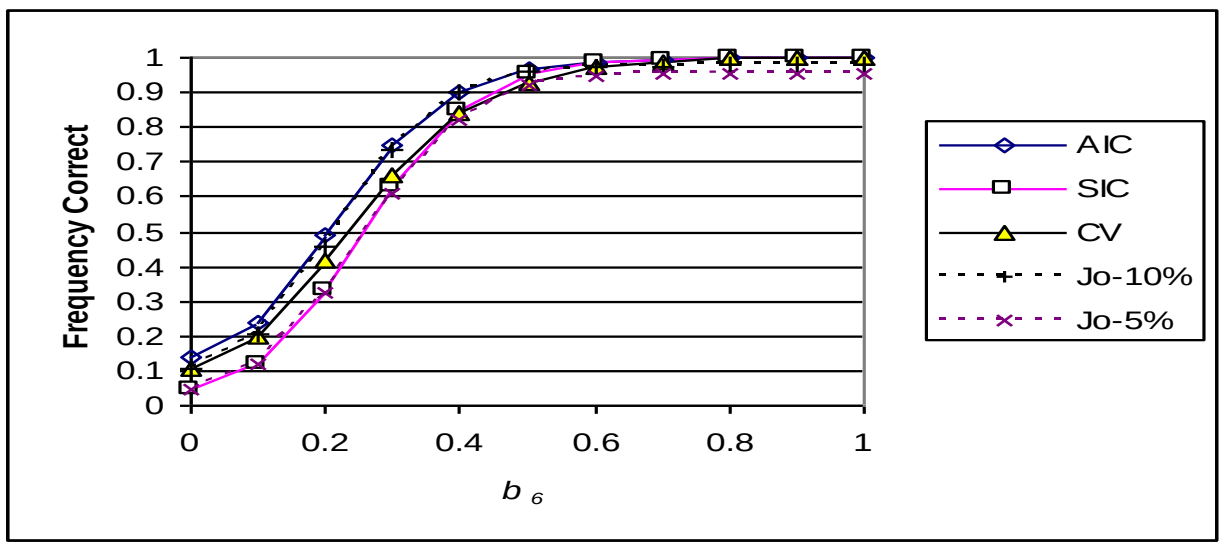

3b: $\Delta Y_t = b_7 \Delta Z_t + u_t$

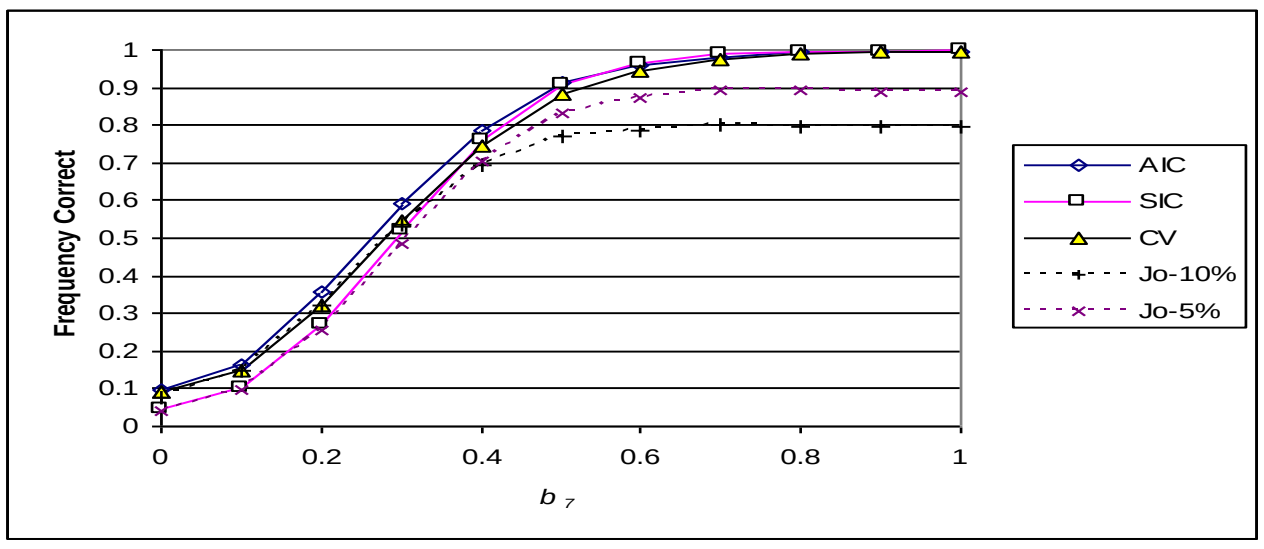

3c: $\Delta Y_t = 0.5\Delta Y_{t-1} + b_8 \Delta Z_{t-1} + u_t$

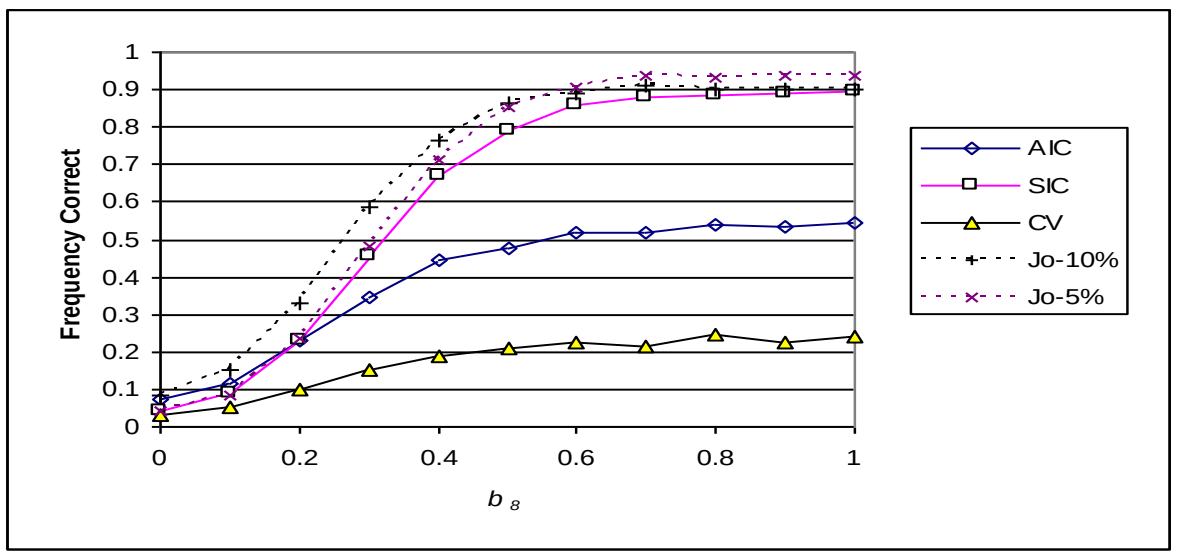

3d: $\Delta Y_t = 0.5\Delta Y_{t-1} + \Delta Z_{t-1} + b_{10}(Y_{t-1} - 1 - Z_{t-1}) + u_t$

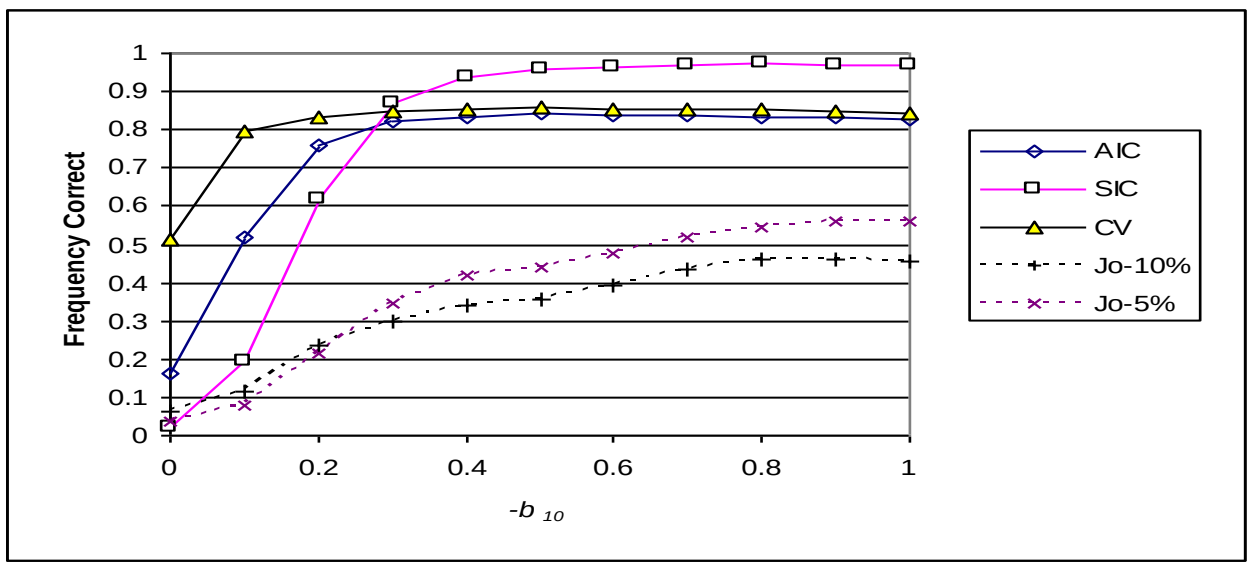

3e: $\Delta Y_t = 1 + b_3 Y_{t-1} + u_t$

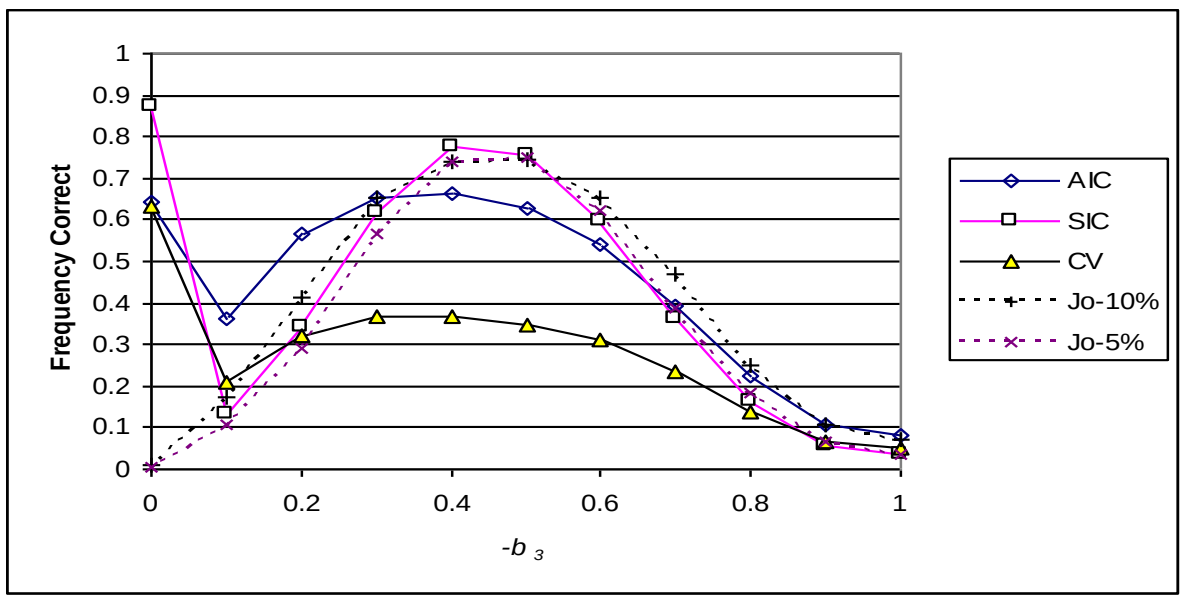

3f: $\Delta Y_t = 0.5\Delta Y_{t-1} + \Delta Z_{t-1} - 0.5(Y_{t-1} - 1 - c_3 Z_{t-1}) + u_t$

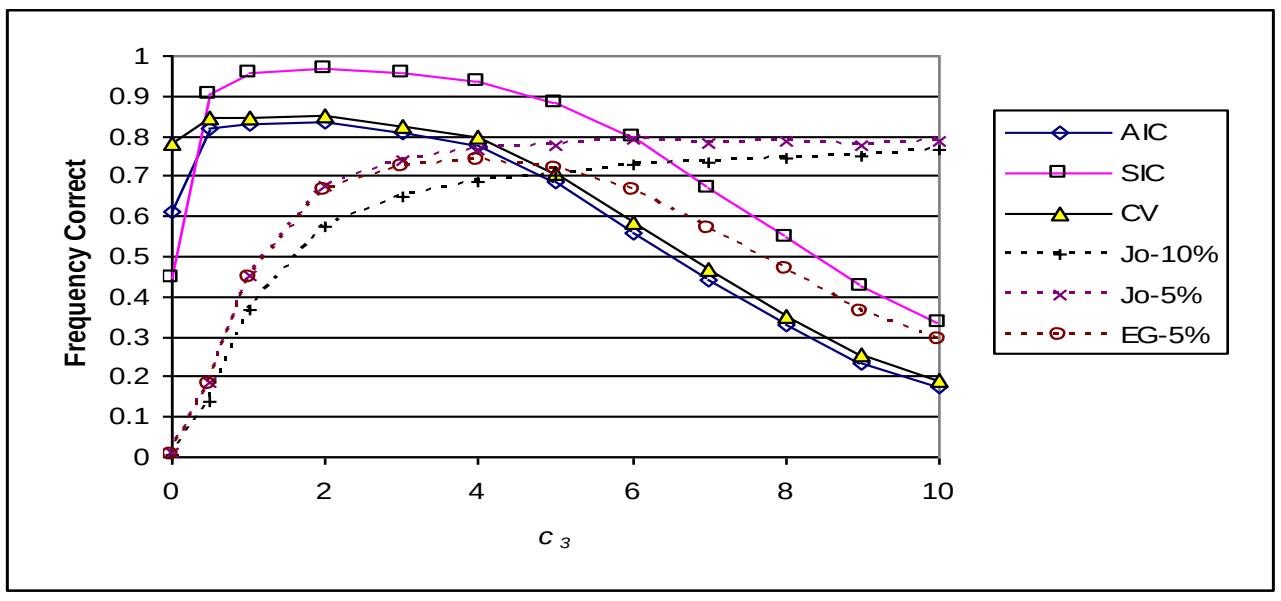

**Figure 3. Frequency of choosing the correct model given various true data generating processes, no trend in $Y$ and $Z$ assumed so models with trend not choosable; $Z_t$ is generated according to $Z_t = Z_{t-1} + \varepsilon_t$ except in case 3a in which it is generated according to $Z_t = 1 + 0.5Z_{t-1} + \varepsilon_t$ ; the varying parameter on horizontal axis is lowest at 0.0001 in all parts and is highest at 0.99999 in 3e.**

Figure 3b considers the situation in which $Y$ is a random walk process related to $Z$ according to the equation $\Delta Y_t = b_7 \Delta Z_t + u_t$, with $Z$ generated by the equation $Z_t = Z_{t-1} + \varepsilon_t$, a random walk. What we see from the response surfaces is that again all the model selection strategies are more successful at choosing the correct model as $b_7$ increases and that the model selection criteria are better performers than the hypothesis-testing strategies above $b_7 = 0.4$. Below that value, all the strategies seem to be close in performance.

Figure 3c deals with $Y$ being related to $Z$ according to the equation $\Delta Y_t = 0.5\Delta Y_{t-1} + b_8\Delta Z_{t-1} + u_t$, with $Z$ generated by a random walk. What we see from the response surfaces is that again SIC is more successful at choosing the correct model as $b_8$ increases, and that SIC performs quite similarly to the hypothesis testing strategies. Notably AIC and CV are considerably worse than SIC at choosing the correct model with $b_8 > 0.2$. This may be attributed to AIC and CV to being less parsimonious in model selection than SIC, so they are more strongly tempted to choose the more complicated cointegration models, which also include the explanatory variables $\Delta Y_{t-1}$ and $\Delta Z_{t-1}$. Note that this was not a problem in Figure 3b, in which the model examined (model 7) had no other competing model that included the same explanatory variable.

Figure 3d deals with a cointegrating relation between $Y$ and $Z$ as the true relation with $Z$ generated as a random walk. The variable $Y$ is generated according to $\Delta Y_t = 0.5\Delta Y_{t-1} + \Delta Z_{t-1} + b_{10}(Y_{t-1} - 1 - Z_{t-1}) + u_t$ and we examine what occurs as $-b_{10}$ varies. We see all the strategies choose the correct model increasingly as $-b_{10}$ increases in magnitude. The hypothesis testing strategies do poorly in choosing the correct model compared to the other strategies almost over the whole range. AIC and especially CV are successful, but by being too accepting of the true model, since when $-b_{10}$ is close to zero they still have high acceptance rates. This different pattern for AIC and CV in comparison to SIC is attributable to these methods being not very parsimonious in model selection in contrast to SIC.

Figure 3e presents a more complicated situation. The variable $Y$ in this case is generated independently from $Z$ such that $Y$ is a stationary process based on the equation $\Delta Y_t = 1 + b_3 Y_{t-1} + u_t$, with $-1 < b_3 < 0$. This figure is interesting because the response surface is dealing with a model straddled between $Y$ being a random walk with drift (if $b_3 = 0$), and $Y$ being a white noise (if $b_3 = -1$; note $-b_3$ is measured on the horizontal axis so the far right is where $b_3 = -1$). In this figure the information criterion and cross-validation strategies have a low frequency of choosing the correct model when the variable on the horizontal axis, $-b_3$, is close to 1 (due to the competing white noise model), have a high frequency in choosing the correct model when $-b_3$ in the middle of the range between 0 and 1, and oddly do well at choosing the correct model when $b_3$ is close to zero. The SIC performance is similar to that of the hypothesis testing strategies except when $b_3$ is close to zero. CV shows an awful performance in this diagram compared to information criteria. The unusual activity of the information criteria and cross-validation when $b_3$ is near zero may be attributed to the fact that we are using prior knowledge that there is no trend in $Y$, so as $b_3$ gets closer to zero the information criteria and the cross-validation measure reject what is apparently looking more like a trend because the trend alternative is not offered as an alternative. When no knowledge of a trend is used, this characteristic for the information criteria and cross validation vanishes (the response surfaces of these strategies then meet the left vertical axis at a more reasonable level of around 5%).

Figure 3f deals with a cointegrating relation between $Y$ and $Z$ as the true relation. The variable $Y$ is generated according to $\Delta Y_t = 0.5\Delta Y_{t-1} + \Delta Z_{t-1} - 0.5(Y_{t-1} - 1 - c_3 Z_{t-1}) + u_t$ with $Z$ generated as a random walk. In this figure we plot the response surface for strategy EG-5% also, unlike in the Figures 3a-3e.[25] The parameter $c_3$, the slope parameter of the cointegrating vector, is allowed to vary from close to zero to a very large value of 10. In this figure, only the hypothesis testing strategies using the Johansen method, Jo-10% and Jo-5%, show the expected pattern of continuously increasing likelihood of choosing the correct model as the parameter $c_3$ increases. The other strategies, including EG-5%, rise then fall in frequency of choosing the correct model as $c_3$ increases.

Notably the hypothesis testing strategies hardly ever choose the model correctly in Figure 3f when $c_3$ is close to zero, which is what we expect to some extent, but AIC, SIC, and CV have substantially large probabilities of choosing cointegration in that neighborhood. The explanation for this oddity is similar to that for the unusual situation in Figure 3e when $-b_3$ was close to zero; when $c_3$ gets closer to zero the process of $Y$ gets closer to getting a trend based on the constant created by the intercept term in the cointegration equation multiplied the -0.5 convergence parameter. Since we are using prior knowledge that there is no trend in $Y$, the information criteria reject what is apparently looking more like a trend because the trend alternative is not offered as an alternative. When no knowledge of a trend is used, this problem for the information criteria vanishes (the response surfaces of AIC and SIC then meet the left vertical axis at a more reasonable levels of around 0.047 and 0.038 respectively).

Figure 4 covers the same situations as Figure 3 in the same order. In this case what is being examined is the frequency of each strategy choosing the correct relation type, i.e. a relation exists between $Y$ and $Z$ in levels, a relation exists between $Y$ and $Z$ only in first differences, a mixed relation, or no relation exists between $Y$ and $Z$. Figures 4a, 4b, 4c, and 4d show patterns similar to their counterparts in Figure 3, with perhaps the most striking differences being CV's notably stronger acceptance of the correct relation in Figure 4a compared to the other strategies for low values of $b_6$ , AIC's higher acceptance of the correct relation for low values of $b_8$ compared to the other strategies in Figure 4c, and the fact that SIC does not outperform AIC and CV for higher values of $-b_{10}$ in Figure 4d.

[25] The response surface for EG-5% is not presented in those previous figures since it would not be substantially different from the response surface for Jo-5% in those other figures, and reducing the number of response surfaces presented helps in the visibility of the remaining response surfaces.

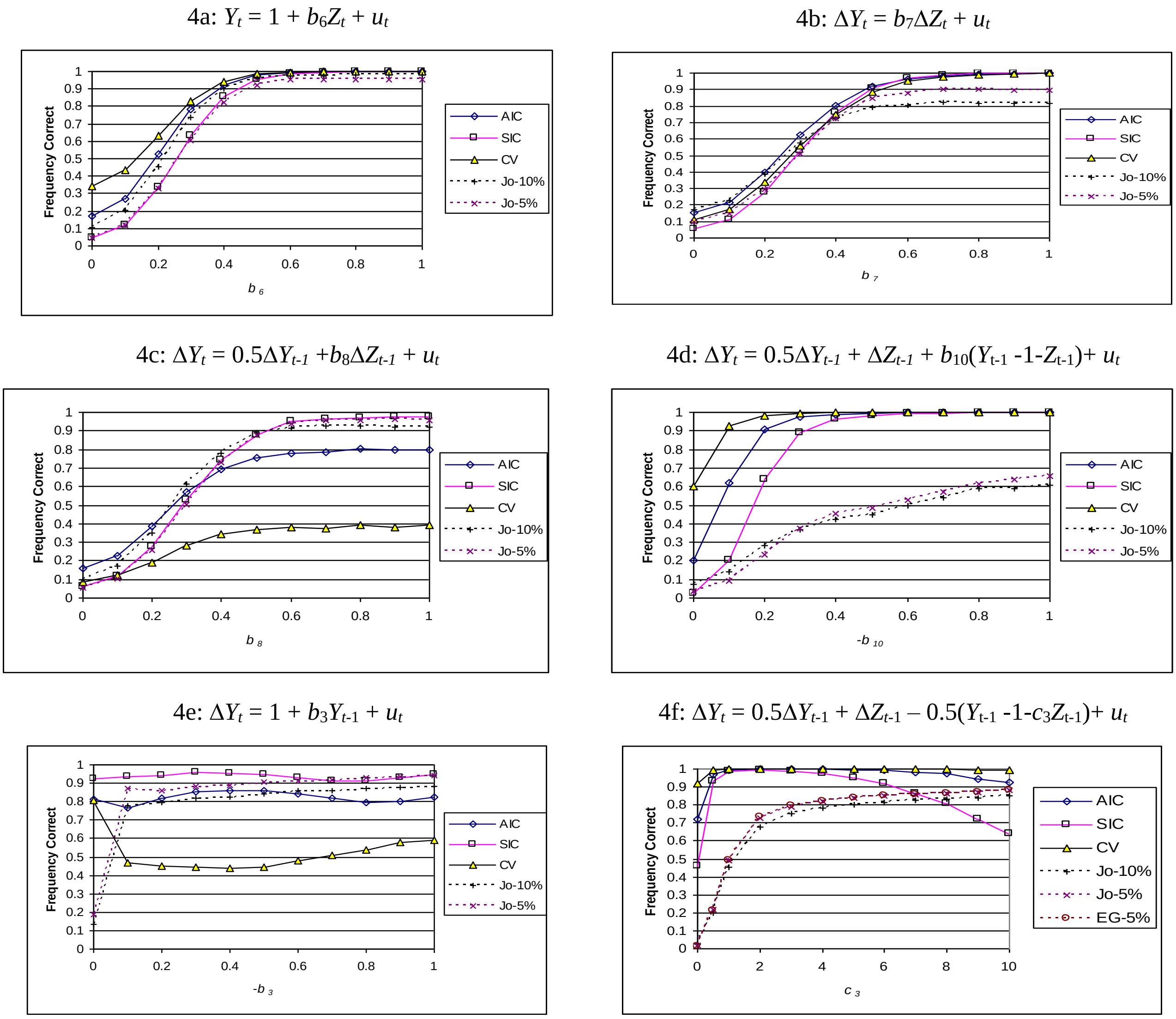

**Figure 4. Frequency of choosing the correct relation type given various true data generating processes, no trends in *Y* and *Z* assumed known so models with trend not choosable; $Z_t$ is generated according to $Z_t = Z_{t-1} + \varepsilon_t$ except in case 4a in which it is generated according to $Z_t = 1 + 0.5Z_{t-1} + \varepsilon_t$ ; the varying parameter on horizontal axis is lowest at 0.0001 in all parts and is highest at 0.99999 in 4e.**

Figure 4e indicates that there is a high acceptance of the true relation type of no relation between *Y* and *Z* by all the strategies over the whole range, except for CV and except when $b_3$ is very close to zero. When $b_3$ gets close to (but not equal to ) zero, which again means getting close to having a trend, then there is a sudden increase in frequency of acceptance of the true relation type by CV and a sudden drop in frequency of acceptance of that true relation type when using the Jo-10% and Jo-5% strategies. The inclusion of white noise as a separate choosable model is important to help the information criteria avoid accepting a relation between $Y$ and $Z$ when some spurious correlation relationships between $Y$ and Z exist between those variables; without it the frequency of choosing the correct relation type in this figure would be substantially lower for SIC and AIC for $-b_3 > 0.5$ (CV would also perform worse).

Figure 4f, dealing with the frequency of choosing the correct relation type, is similar to Figure 3f which uses the same simulations, but it is notable that the response surface from EG-5% is now closely following that for Jo-5%. AIC, SIC, and CV show notably better performances at choosing the correct relation type (compared to their performances in choosing the correct model) at the high levels of $c_3$, but similar to what we see in Figure 3f, there is a decline in the acceptance of the true relation type as $c_3$ increases at the high levels (although this is hardly perceptible for CV over the values of $c_3$ given).

Figure 5 repeats the situation in Figures 3f and 4f except the equation generating $\Delta Y$ is $\Delta Y_t = 1 + 0.5\Delta Y_{t\text{-}1} + \Delta Z_{t\text{-}1} - 0.5(Y_{t\text{-}1} - 1 - c_3 Z_{t\text{-}1}) + u_t$ rather than the equation $\Delta Y_t = 0.5\Delta Y_{t\text{-}1} + \Delta Z_{t\text{-}1} - 0.5(Y_{t\text{-}1} - 1 - c_3 Z_{t\text{-}1}) + u_t$, $Z$ has a random walk with drift, and the trend status is considered unknown *a priori*. In this situation a very different pattern emerges. As $c_3$ increases from zero, the information criteria still rise and fall in choosing the correct model, but the hypothesis-testing strategies seem hardly ever to choose the correct model. The hypothesis testing strategies do choose the correct relation type more often as $c_3$ increases, but now the information criteria choose the correct relation type substantially more often than the hypothesis testing strategies for $0.1 \le c_3 \le 10$.

5a: frequency Choosing Correct Model

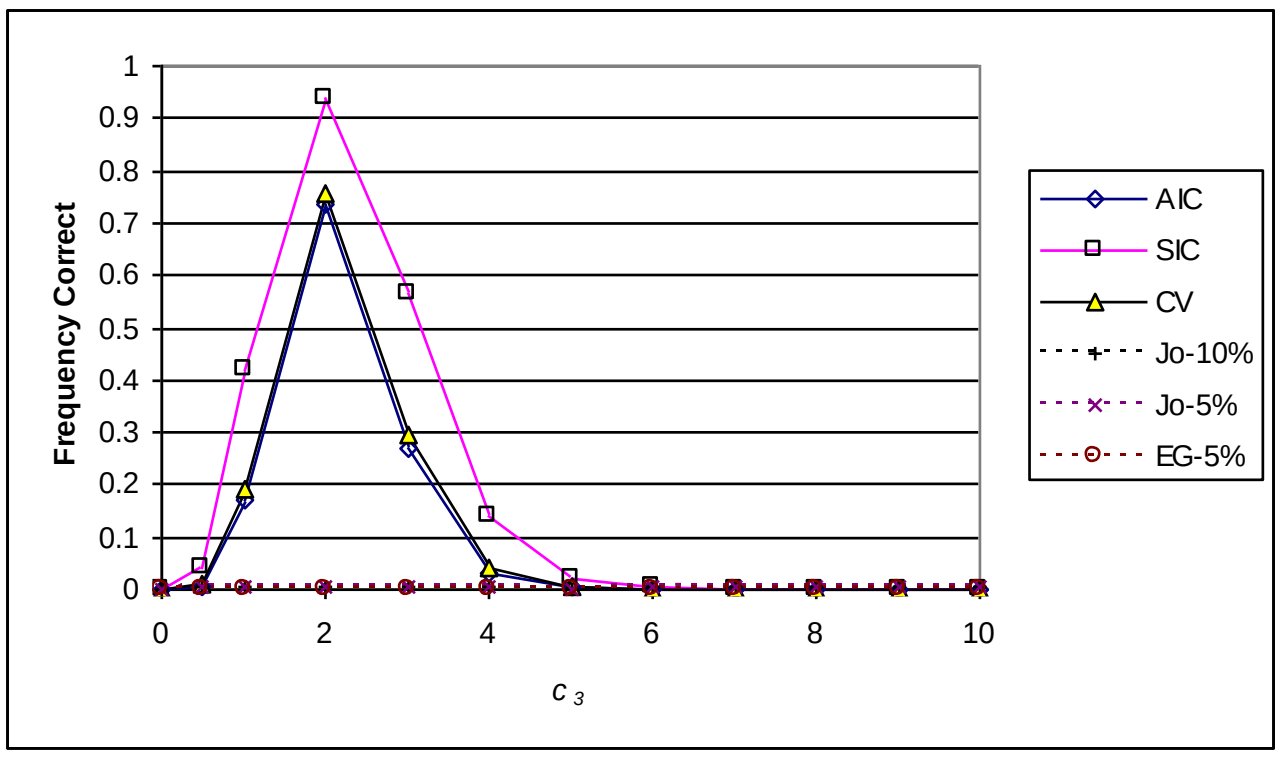

5b: Frequency Choosing Correct Relation Type

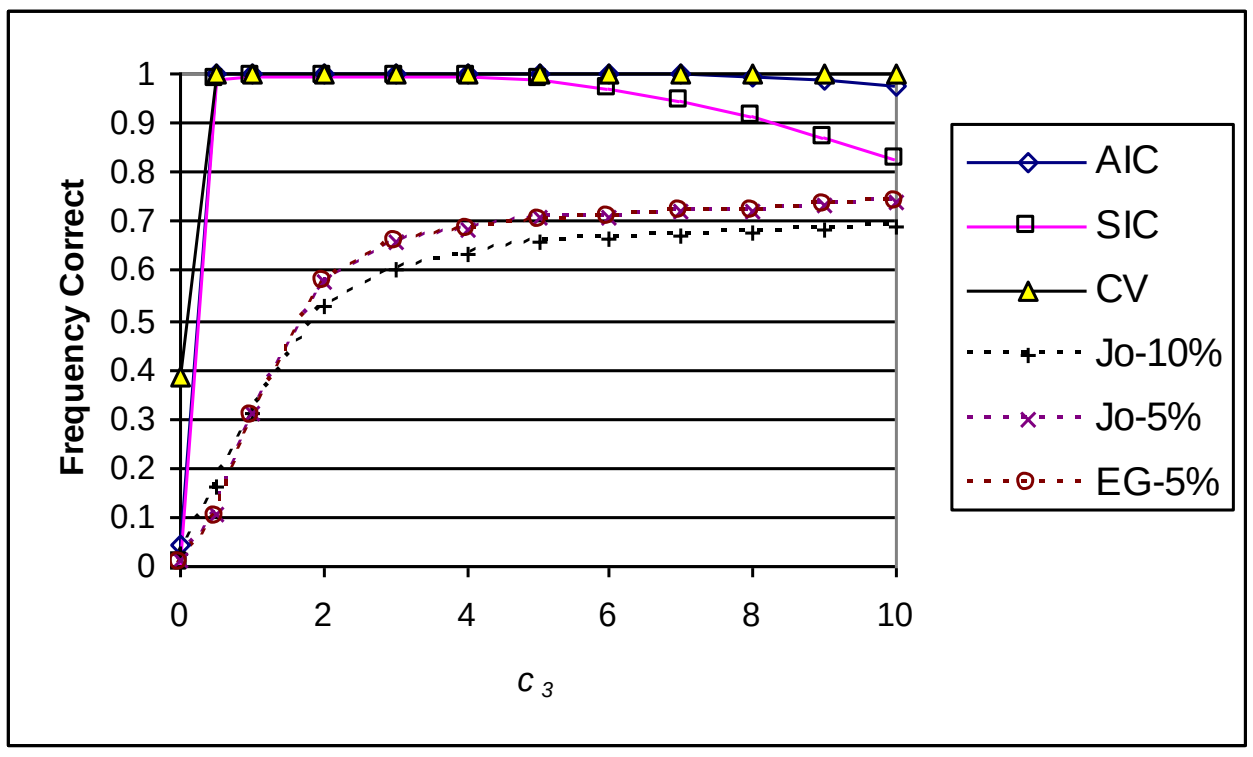

**Figure 5. Frequencies of choosing the correct model and choosing the correct relation type for the data generating process $\Delta Y_t = 1 + 0.5\Delta Y_{t\text{-}1} + \Delta Z_{t\text{-}1} - 0.5(Y_{t\text{-}1} - 1 - c_3 Z_{t\text{-}1}) + u_t$ with an unknown trend status for $Y$ and $Z$; $Z_t$ generated according to $Z_t = 1 + Z_{t\text{-}1} + \varepsilon_t$ ; $c_3$ lowest at 0.0001.**

One of the difficulties with the performance measures examined in Figures 3–5 is that choosing a model or relation type correctly may be not what we ideally want the strategy to do when the true data generating process is getting close to another model or relation type respectively. For example, in Figures 3d and 4d the performance of AIC and CV at choosing the correct model or relation type when $b_{10}$ is not equal to zero but close to it seems wonderful compared to the other strategies (except for the implication for choosing the correct model or relation when $b_{10} = 0$), but the other strategies probably choose models with better predictive capability when $b_{10}$ is close to zero since they are less frequently trying to estimate $b_{10}$ under those circumstances; constraining that parameter to be zero would tend to provide better predictions than relying upon estimates of it.

Measuring performance based on the ability to choose a model or relation type correctly also has the drawback that the performance jumps when a parameter goes from zero to slightly not zero. Those strategies that do best at choosing a model or relation type correctly when a parameter is zero are exactly those which do worst at that when that parameter is slightly not zero (assuming the parameter being zero or not distinguishes different models or relation types). It is odd to have a measure of performance that is so sensitive to such a slight difference.

It seems also that measuring performance on the ability of a strategy to choose the correct model or relation type (a class of models) is unusual to apply in the social sciences since it is simply incredible to consider that the correct model is among the models we are considering. All our models in the social sciences are simply approximations to a far more complex reality. Under such circumstances it seems more natural to consider model-choosing success as more frequently choosing those models that have better predictive performance.

Due to the above arguments, we consider the $L_2$ distance measure described below for evaluating performance. For simulation $s$ the $L_2$ distance for $\Delta Y$ is given by

$$L_{2s} = \frac{\sum_t \left(E(\Delta Y_{ts} \mid \chi_t) - \Delta \hat{Y}_{ts}\right)^2}{T}, \tag{15}$$

where $\chi_t$ is the vector of explanatory variable values at time $t$, and $\Delta\hat{Y}_{ts}$ is the estimated value of $\Delta Y_t$ in simulation $s$ given the estimated equation associated with a particular model and $\chi_t$.[26] The term $E(\Delta Y_{ts} \mid \chi_t)$ is the expected value of $\Delta Y_t$ given $\chi_t$ and given we know the true parameters. $L_{2s}$ is thus simply the average squared difference between expected and estimated values for the dependent variable (McQuarrie and Tsai, 1998), $\Delta Y$, in simulation $s$. A lower $L_{2s}$ distance indicates better predictive performance. For our simulations we take the mean of $L_{2s}$ over the simulations for a given set of parameters and refer to that mean as $L_2$.

Parts a-f of Figure 6 cover the same situations as parts a-f in Figures 3 and 4 in the same order, but ln $L_2$ is instead used as the performance measure. The natural log of the $L_2$ distance is used since there are in some cases rather large differences in $L_2$ between strategies, so a rescaling is needed to better observe the differences in the patterns of $L_2$. The differences between ln $L_2$ values for different strategies represents the log of the ratio of the underlying $L_2$ values. For example, since when $b_6$ =0.5 in Figure 6a the ln $L_2$ value for SIC is about -3 and the ln $L_2$ value for Jo-5% is about -2.75, then that implies the $L_2$ for Jo-5% is about 28% higher than the $L_2$ for SIC at that point ( $\exp(-1.5-(-2.75)-1)=0.28=28\%$ ).

What is notable about these diagrams is the exceptionally good predictive performance of SIC compared to the hypothesis testing strategies in all six cases. That is true even in Figure 6f with high values of $c_3$. If one recalls from the previous figures, that was the situation (shown in Figures 3f and 4f) in which SIC performed worst in choosing the model or relation compared to the hypothesis-testing strategies when knowledge of no trends in $Y$ and $Z$ was assumed. Even though SIC performs badly at choosing the correct model of cointegration at high values of $c_3$ it is doing so for a good reason apparently—it is finding one or more competing models that have better predictive power. In this case the primary models it most frequently chooses as alternatives are models 9.02 (difference Granger-causal model, no intercept, two augmentation lags) and 13.02 (error correction model, no intercept, two augmentation lags). It is interesting to note that what seems to be very similar patterns of performance between SIC and the hypothesis-testing strategies in Figure 3a, Figure 3b, and the right half of Figure 3e results in such different patterns of predictive performance in the corresponding Figures 6a, 6b, and 6e.

[26] There exists a broader class of $L_p$ distance measures that takes the $p$th power of the absolute difference between various pairs of values and find the mean over the resulting numbers. $L_2$ simply uses $p$ = 2. This measure is closely associated with, but not exactly same as the predictive mean square error, $\sum_t E\left((\Delta Y_{ts} \mid \chi_{ts}) - \Delta\hat{Y}_{ts}\right)^2 / T$.

6a: $Y_t = 1 + b_6 Z_t + u_t$

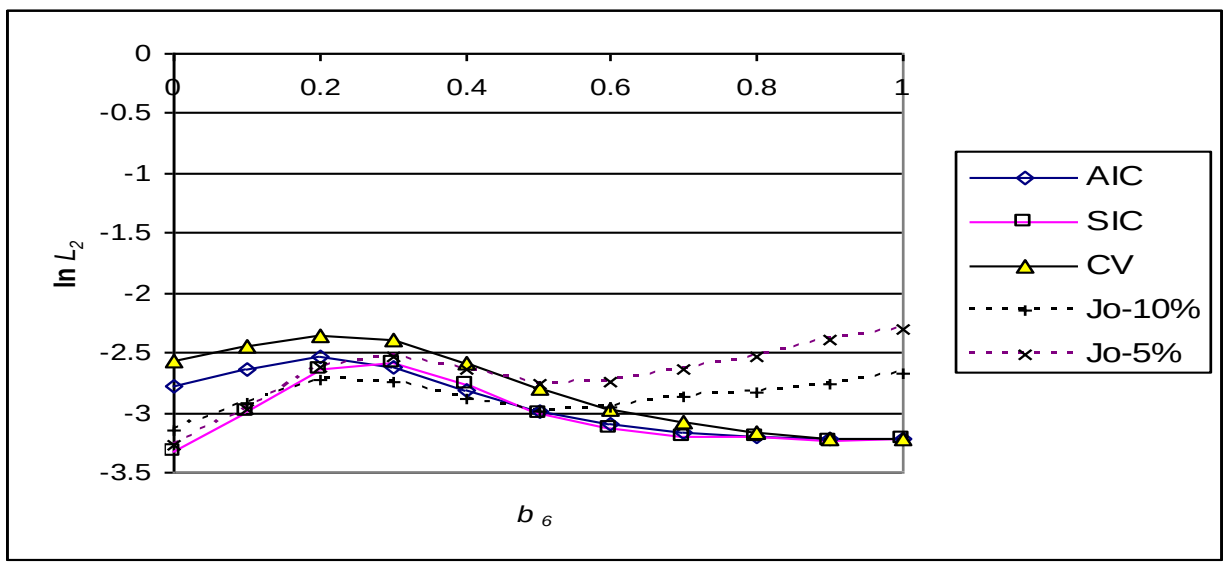

6b: $\Delta Y_t = b_7 \Delta Z_t + u_t$

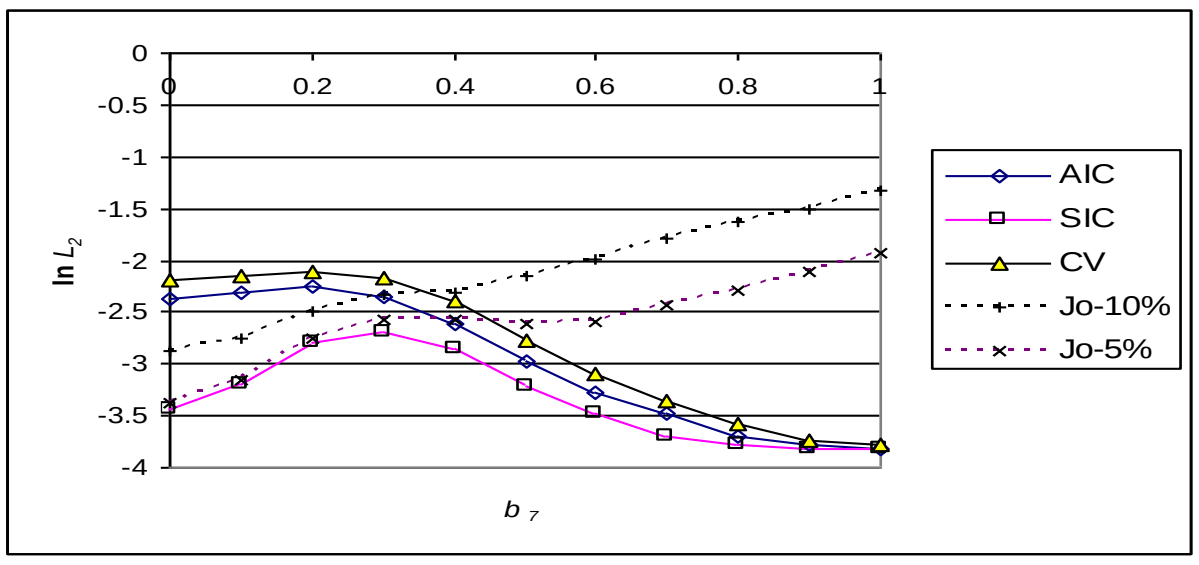

6c: $\Delta Y_t = 0.5\Delta Y_{t-1} + b_8 \Delta Z_{t-1} + u_t$

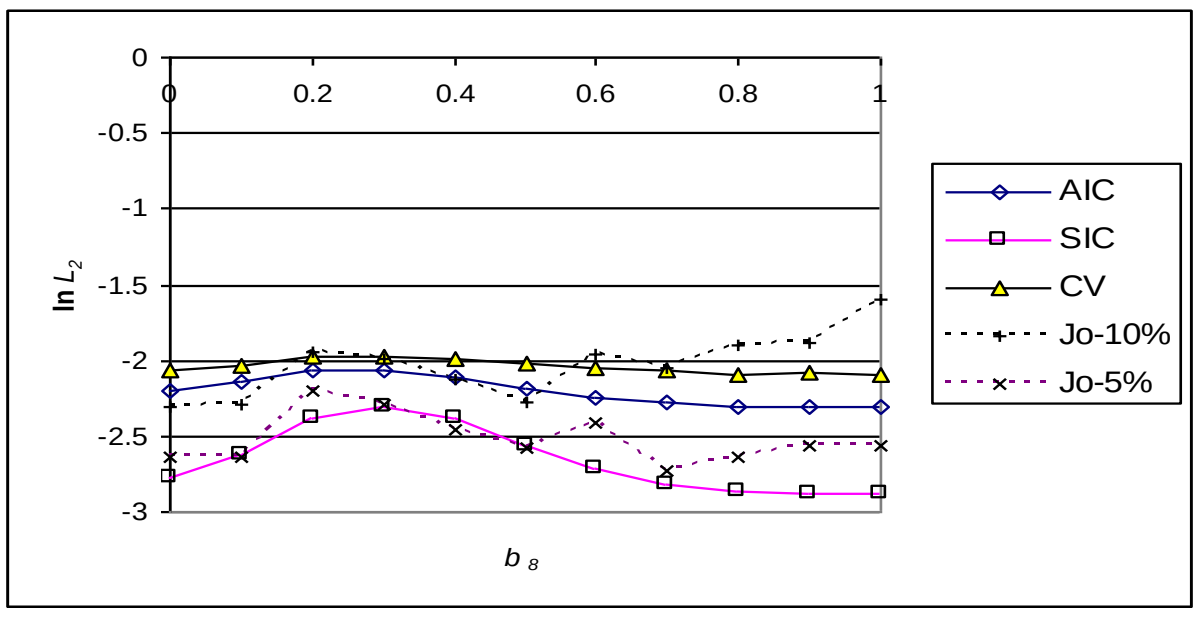

6d: $\Delta Y_t = 0.5\Delta Y_{t-1} + \Delta Z_{t-1} + b_{10}(Y_{t-1} - 1 - Z_{t-1}) + u_t$

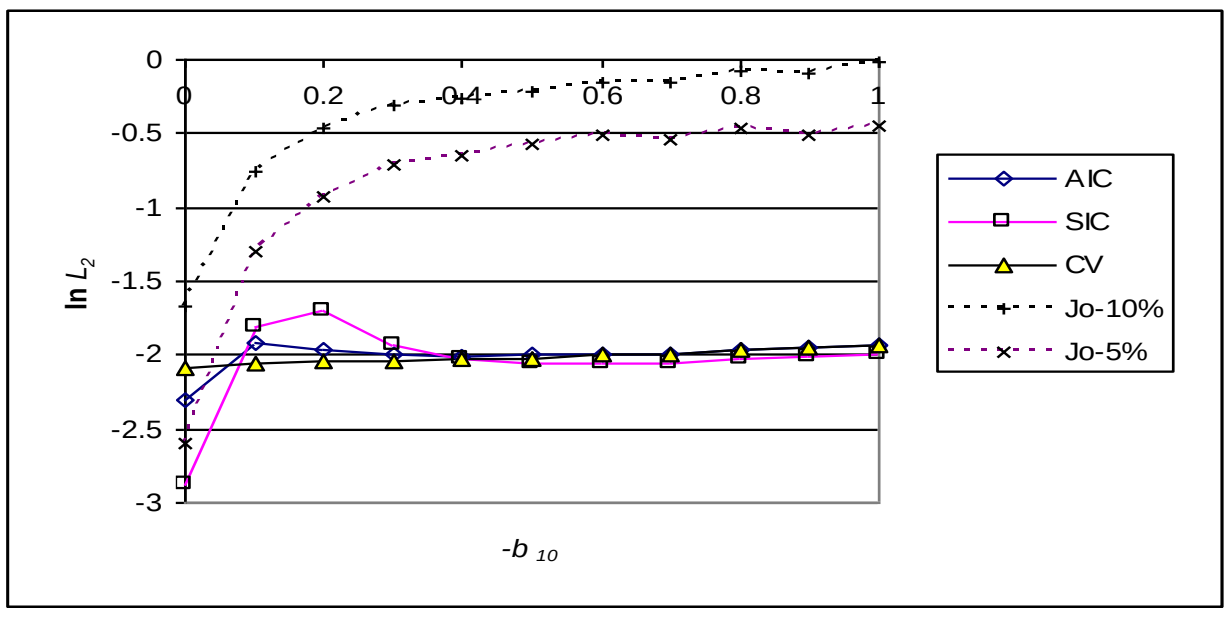

6e: $\Delta Y_t = 1 + b_3 Y_{t-1} + u_t$

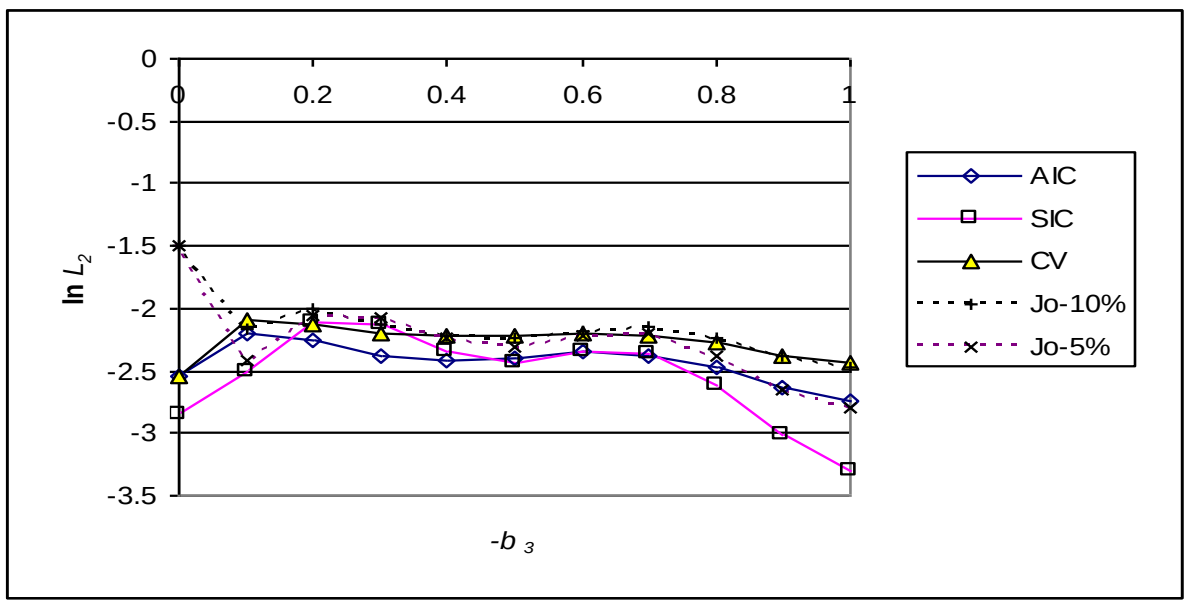

6f: $\Delta Y_t = 0.5\Delta Y_{t-1} + \Delta Z_{t-1} - 0.5(Y_{t-1} - 1 - c_3 Z_{t-1}) + u_t$

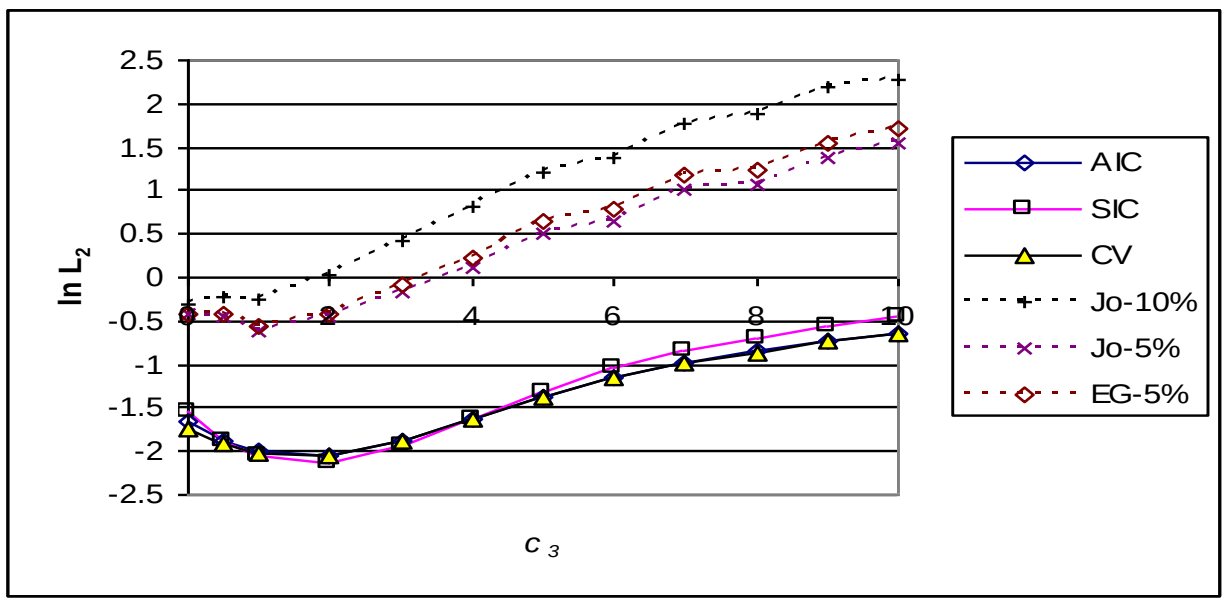

**Figure 6. ln $L_2$ for various true data generating processes, no trend assumed known so models with trend not choosable; $Z_t$ is generated according to $Z_t = Z_{t-1} + \varepsilon_t$ except in case 6a in which it is generated according to $Z_t = 1 + 0.5Z_{t-1} + \varepsilon_t$**

What is also notable about these diagrams is that SIC performs better than or about as well as AIC and CV in all six cases over almost all parameter values. One of the notable exceptions is that AIC and CV have somewhat better $L_2$ performance for some low-magnitude $-b_{10}$ values in Figure 6d. Not surprisingly, AIC and CV pay for their previous poor performances in Figures 3c and 4c with substantially poorer predictive performances compared to SIC and Jo-5% in Figure 6c. Unusually Jo-10% has similarly poor predictive performance in Figure 6c, despite the good performance shown for this strategy in Figures 3c and 4c.

We have re-run the simulations producing Figures 6 to deal with the situations where the researcher knows there is a trend in both variables or the researcher does not know the trend status. Under these circumstances we had $Z$ following a random walk with drift and included an additional intercept of 1 in the ΔY generating processes in the cases of Figures 6b, 6c, 6d, and 6f (so no relation to *Z*, Δ*Z*, or lagged Δ*Z* would still result in *Y* having a trend in these cases). In those simulations we found similarly superior $L_2$ properties for SIC compared to the other strategies and otherwise the most interesting cases of changes in patterns are along the lines noted in the discussion with Figures 3e and 3f.

## 7. Minimizing the maximum regret

This section investigates a much wider variety of true generating processes than presented in the last section. As may be expected, one true data generating process (including a specific set of parameter values), will favor one of the model selection strategies as being optimal whereas another true data generating process will favor another. Since the true data generating process is unknown to the researcher in practice, we use the principal of minimax regret over various parameter permutations as a way of evaluating the relative performance of the various strategies.[27] Supposing that $G_{\theta k}$ is a measurement of goodness of a model (frequency of choosing the correct model, frequency of choosing the correct relation type, or -ln $L_2$) for a given parameter permutation θ when using model-choosing strategy *k*, then the regret of choosing strategy *k* rather than another strategy $k'$ (not *k*) for the parameter permutation would be $G_{\theta k'} - G_{\theta k}$. Between two competing method-choosing strategies $\Psi_1$ and $\Psi_2$, the minimax regret strategy would be

$$\arg\min_{k\in(\psi_1,\psi_2)}(\max_{\theta\in\Theta}(G_{\theta k'} - G_{\theta k})) \ , \qquad (16)$$

where $k' \in (i, j)$ and $\Theta$ is the set of all considered parameter permutations. Before presenting the minimax regret results for different regret measurements, we will describe the various parameter permutations we will use for the true model.

For the true model generating *Y* based on equation (1) and generating *Z* based on equation (10), the assorted permutations of values from the following sets are used:

$b_1$ = {0, 1}, $b_2$ = {0, 0.5, 1}, $b_3$ = {-1, -0.9, -0.5, -0.1, 0}, $b_4$ = {0, 0.5}, $b_5$ = {0, 0.3},
$b_6$ = {0, 0.1, 1, 10}, $b_7$ = {0, 0.1, 1, 10}, $b_8$ = {-0.5, 0, 0.1, 0.5, 1, 10}, $b_9$ = {0, 0.1, 1, 10},
$b_{10}$ = {0, -0.1, -0.5, -0.8, -1}, $c_1$ = {0, 1}, $c_2$ = {0, 0.1, 1, 10},

[27] Hacker (2010) uses a similar strategy of considering minimax regret for evaluating the performance of information criteria and hypothesis-testing strategies in determining unit-root status and trend status for a single variable.

$m_1 = \{0, 1\}$, $m_2 = \{0, 1\}$, $m_3 = \{1, 0.5\}$.[28]

There are 14,745,600 permutations of these parameter values, but not all of those permutations are used since we exclude permutations that do not match one of the models listed in Table 1 (each model in that table is identified by an associated equation with nonzero parameters). When $b_{10} = 0$, for example, then permutations with $c_1$ or $c_2$ nonzero are not used and when $b_{10} \neq 0$, then only permutations in which only $c_1$ and $c_2$ both nonzero are used. As another example, a parameter permutation with $b_6$ nonzero and $b_7$ nonzero is not used since there is no model in Table 1 that has that permutation. We also exclude the following: (I) permutations that result in model 15 or 16,[29] (II) permutations that involve $Z$ not following one of the processes listed in Table 2, (III) permutations in which $Z$ follows a random walk (with or without drift) and the $b_i$ ($i$ = 1,…10) parameters results in model 11 or 12 (current level relations without or with trend), and (IV) permutations in which $Z$ is a stationary process (with or without trend) and the $b_i$ parameters results in model 13 or 14 (the error correction models). After the exclusions we have 1090 valid permutations of parameter values used for our simulations overall.

When we make the assumption that the researcher knows that both variables have no trend, we also exclude those permutations of $b_i$ that would result in one of the even-numbered models for $Y$'s process (models 2, 4, 6, etc) and those permutations of $m_1$, $m_2$, and $m_3$ parameters that would result in a trend for $Z$. That results in 259 valid permutations of parameter values. When we make the assumption that the researcher knows that each of the variables has a trend, we also exclude those permutations of $b_i$ that would result in one of the odd-numbered models for $Y$'s process (models 1, 3, 5, etc) and those permutations of $m_1$, $m_2$, and $m_3$ parameters that would result in no trend for $Z$. That results in 286 valid permutations of parameter values.

The $b_3$ values are chosen to be in the range [-1, 0] to consider the unit root situation and various nonoscillatory stationary processes for $Y_t$ when $b_4 = b_5 = \ldots = b_{10} = 0$. The $b_{10}$ values are chosen to be in the range [-1, 0] to consider cointegration situations in which the speed of convergence to equilibrium is very fast ($b_{10}$=-1) to very slow and to consider no-cointegration situations ($b_{10}$=0). The superior performance of CV and AIC over SIC in choosing the correct model or relation type when there is a slow speed of convergence, as seen in the left part of Figures 3d and 4d, is thus included in the results of these simulations. The poor performance of information criteria relative to Jo-5% and Jo-10% in choosing the correct model for high values of $c_3$, as seen in Figure 3f, is covered by having a high value for $c_3$ (a value of 10) as a possibility in these simulations.

[28] For purely programming purposes there are some rather benign variations to this list: when $b_{10}$ is listed as -1, it is actually -0.99999, and when dealing with model 13, i.e. cointegration models, $b_1 = 0$ and $c_1 = 0$ are in actuality respectively $b_1 = 0.00001$ $c_1 = 0.00001$.

[29] These models are excluded as true models since they cannot be chosen by the hypothesis choosing strategies, as seen in Figure 1.

The parameters $b_6$, $b_7$, $b_8$ and $b_9$ are chosen with a wide variety of values—including zero, a small value, and a large value—since a relation between $Y$ (or $\Delta Y$) and $Z$ or changes in $Z$ is a major focus in this paper. The parameter $b_8$ is allowed to take on a wider variety of values to provide richer consideration of how sensitivity of $\Delta Y_t$ to $\Delta Z_{t-1}$ affects the ability of the various strategies in finding $\Delta Y_t$ is related to $Z_{t-1}$ through the cointegrating vector. The maximum values for $b_4$ and $b_5$ are chosen so the sum of the coefficients for $\Delta Y_{t-1}$ and $\Delta Y_{t-2}$ sum to less than 1, thereby avoiding $\Delta Y_t$ having a unit root.

The strategies we consider in this section are broader than in the previous section. We consider AIC, AICc AICu and SIC among information criteria strategies, along with CV and six hypothesis-testing strategies: EG-10%, EG-5%, EG-10/5, Jo-10%, J0-5%, and Jo-10/5.
We measure regret in three different ways. The first is how much more frequently the wrong model is chosen compared to another procedure. Table 3 provides maximum regrets measured in this fashion over all the simulations when it is assumed that the researcher knows there is no trend. Each cell's value shows the maximum regret (highest increased frequency of choosing the wrong model) when using the procedure listed for the cell's row rather than using the procedure listed for the cell's column. For example, when using EG-5% rather than AIC, the maximum increased frequency of choosing the wrong model over all the simulations of various parameter permutations is 94 percentage points, whereas when using AIC rather than EG-5%, the maximum increased frequency of choosing the wrong model over all the simulations of various parameter permutations is 44 percentage points. Using the concept of minimax regret, AIC is more favorable to use in comparison to EG-5% since AIC has the lower maximum regret compared to EG-5%. The table shows that SIC minimizes the maximum regret when compared to each of the other strategies. The shaded cells are those used in supporting that statement: 0.40 < 0.47 (comparing SIC with AIC), 0.28 < 0.40 (comparing SIC with AICc), and so forth.

There is a notable difference in this table between the performance of the information criteria and cross validation on one hand versus hypothesis testing strategies on the other. The maximum regrets of using any of the hypothesis testing strategies instead of any of the information criteria or cross validation are always at least 87 percentage points, while the corresponding maximum regrets of any of the information criteria or cross validation against any of the hypothesis testing strategies are always lower.

**Table 3. Maximum regret on frequency of choosing correct model, no trend in *Y* and *Z* in data generating process, models with trends in these variables not choosable**[a]

| | AIC | AICc | AICu | SIC | CV | EG-10% | EG-5% | EG-10/5 | Jo-10% | Jo-5% | Jo-10/5 |
|---|---|---|---|---|---|---|---|---|---|---|---|
| AIC | 0.00 | 0.14 | 0.36 | 0.47 | 0.49 | 0.38 | 0.44 | 0.40 | 0.78 | 0.78 | 0.42 |
| AICc | 0.17 | 0.00 | 0.28 | 0.40 | 0.56 | 0.26 | 0.37 | 0.33 | 0.75 | 0.76 | 0.35 |
| AICu | 0.38 | 0.22 | 0.00 | 0.12 | 0.63 | 0.09 | 0.11 | 0.10 | 0.70 | 0.71 | 0.11 |
| SIC | 0.40 | 0.28 | 0.10 | 0.00 | 0.66 | 0.10 | 0.09 | 0.11 | 0.70 | 0.71 | 0.11 |
| CV | 0.55 | 0.64 | 0.71 | 0.71 | 0.00 | 0.65 | 0.68 | 0.65 | 0.77 | 0.78 | 0.71 |
| EG-10% | 0.90 | 0.88 | 0.87 | 0.87 | 0.96 | 0.00 | 0.14 | 0.13 | 0.72 | 0.73 | 0.66 |
| EG-5% | 0.94 | 0.93 | 0.90 | 0.90 | 0.96 | 0.14 | 0.00 | 0.16 | 0.72 | 0.72 | 0.54 |
| EG-10/5 | 0.95 | 0.93 | 0.90 | 0.90 | 0.96 | 0.10 | 0.14 | 0.00 | 0.72 | 0.73 | 0.53 |
| Jo-10% | 0.89 | 0.88 | 0.87 | 0.87 | 0.96 | 0.16 | 0.23 | 0.16 | 0.00 | 0.14 | 0.17 |
| Jo-5% | 0.89 | 0.88 | 0.87 | 0.87 | 0.96 | 0.14 | 0.17 | 0.16 | 0.14 | 0.00 | 0.16 |
| Jo-10/5 | 1.00 | 1.00 | 1.00 | 1.00 | 1.00 | 0.76 | 0.82 | 0.75 | 0.83 | 0.84 | 0.00 |

[a] Each cell's value shows the maximum regret (highest increased frequency of choosing the wrong model) when using the procedure listed for the cell's row rather than using the procedure listed for the cell's column. The shaded cells are those associated with the strategy that has minimax regret compared to all other strategies.

Table 4 presents maximum regrets in which regrets are measured as how much more frequently the wrong relation type is chosen compared to another procedure, using prior knowledge of no trend in *Y* and *Z*. Otherwise the table is developed in the same way as Table 3. The table shows that AIC minimizes the maximum regret when compared to each of the other strategies, with the shaded cells being those used in supporting that statement. As in Table 3, the maximum regrets of any of the hypothesis testing strategies are high against any of the information criteria or cross validation, while the corresponding maximum regrets of any of the information criteria or cross validation against any of the hypothesis testing techniques are always lower.

**Table 4. Maximum regret on frequency of choosing correct relation type, no trend in *Y* and *Z* in data generating process, models with trends in these variables not choosable**[a]

| | AIC | AICc | AICu | SIC | CV | EG-10% | EG-5% | EG-10/5 | Jo-10% | Jo-5% | Jo-10/5 |
|---|---|---|---|---|---|---|---|---|---|---|---|
| AIC | 0.00 | 0.14 | 0.27 | 0.28 | 0.56 | 0.36 | 0.38 | 0.38 | 0.50 | 0.51 | 0.39 |
| AICc | 0.18 | 0.00 | 0.14 | 0.17 | 0.65 | 0.27 | 0.30 | 0.27 | 0.50 | 0.51 | 0.25 |
| AICu | 0.42 | 0.27 | 0.00 | 0.03 | 0.76 | 0.50 | 0.53 | 0.50 | 0.63 | 0.55 | 0.14 |
| SIC | 0.47 | 0.32 | 0.06 | 0.00 | 0.79 | 0.52 | 0.54 | 0.51 | 0.67 | 0.58 | 0.15 |
| CV | 0.64 | 0.70 | 0.76 | 0.77 | 0.00 | 0.64 | 0.71 | 0.71 | 0.64 | 0.71 | 0.71 |
| EG-10% | 0.90 | 0.89 | 0.87 | 0.87 | 0.94 | 0.00 | 0.15 | 0.13 | 0.67 | 0.67 | 0.68 |
| EG-5% | 0.94 | 0.93 | 0.92 | 0.91 | 0.95 | 0.09 | 0.00 | 0.02 | 0.71 | 0.70 | 0.55 |
| EG-10/5 | 0.94 | 0.93 | 0.91 | 0.91 | 0.95 | 0.10 | 0.15 | 0.00 | 0.70 | 0.69 | 0.55 |
| Jo-10% | 0.86 | 0.85 | 0.84 | 0.84 | 0.94 | 0.13 | 0.15 | 0.15 | 0.00 | 0.15 | 0.17 |
| Jo-5% | 0.88 | 0.87 | 0.86 | 0.86 | 0.95 | 0.10 | 0.09 | 0.09 | 0.14 | 0.00 | 0.11 |
| Jo-10/5 | 0.97 | 0.97 | 0.96 | 0.96 | 0.98 | 0.84 | 0.86 | 0.83 | 0.85 | 0.88 | 0.00 |

[a] Each cell's value shows the maximum regret (highest increased frequency of choosing the wrong relation type) when using the procedure listed for the cell's row rather than using the procedure listed for the cell's column. The shaded cells are those associated with the strategy that has minimax regret compared to all other strategies.

**Table 5. Maximum regret on ln $L_2$, no trend in $Y$ and $Z$ in data generating process, models with trends in these variables not choosable[a]**

| | AIC | AICc | AICu | SIC | CV | EG-10% | EG-5% | EG-10/5 | Jo-10% | Jo-5% | Jo-10/5 |
|---|---|---|---|---|---|---|---|---|---|---|---|
| AIC | 0.00 | 0.16 | 0.71 | 1.09 | 0.18 | 0.42 | 0.94 | 0.64 | 0.51 | 1.00 | 0.80 |
| AICc | 0.13 | 0.00 | 0.56 | 0.94 | 0.22 | 0.27 | 0.79 | 0.51 | 0.36 | 0.85 | 0.65 |
| AICu | 0.36 | 0.23 | 0.00 | 0.38 | 0.44 | 0.01 | 0.23 | 0.25 | 0.02 | 0.28 | 0.25 |
| SIC | 0.42 | 0.35 | 0.12 | 0.00 | 0.50 | 0.00 | 0.01 | 0.11 | 0.00 | 0.01 | 0.12 |
| CV | 0.29 | 0.43 | 0.88 | 1.25 | 0.00 | 0.59 | 1.11 | 0.86 | 0.67 | 1.16 | 0.97 |
| EG-10% | 8.20 | 8.24 | 8.29 | 8.29 | 8.13 | 0.00 | 2.48 | 2.14 | 2.24 | 2.75 | 2.59 |
| EG-5% | 7.34 | 7.38 | 7.43 | 7.43 | 7.27 | 0.72 | 0.00 | 0.72 | 0.72 | 1.23 | 1.05 |
| EG-10/5 | 7.95 | 7.99 | 8.03 | 8.04 | 7.89 | 0.18 | 1.64 | 0.00 | 1.37 | 1.86 | 1.53 |
| Jo-10% | 7.96 | 8.00 | 8.05 | 8.05 | 7.88 | 0.16 | 0.98 | 0.84 | 0.00 | 1.12 | 0.88 |
| Jo-5% | 7.11 | 7.15 | 7.20 | 7.21 | 7.04 | 0.72 | 0.13 | 0.72 | 0.71 | 0.00 | 1.04 |
| Jo-10/5 | 7.61 | 7.66 | 7.70 | 7.71 | 7.54 | 0.18 | 0.71 | 0.00 | 0.19 | 0.89 | 0.00 |

[a] Each cell's value shows the maximum regret (greatest increase in the ln $L_2$ distance) when using the procedure listed for the cell's row rather than using the procedure listed for the cell's column. The shaded cells are those associated with the strategy that has minimax regret compared to all other strategies.

Table 5 presents maximum regrets when regret is measured as how much higher the ln $L_2$ distance is compared to that when using another procedure, using again prior knowledge of no trend in $Y$ and $Z$. Otherwise the table is developed in the same way as Tables 3 and 4. The table shows that SIC minimizes the maximum regret when compared to each of the other strategies, with the shaded cells being those used in supporting that statement. As in the previous two tables, the maximum regrets of any of the hypothesis testing strategies are higher than those of any of the information criteria or cross validation, while the corresponding maximum regrets of any of the information criteria or cross validation against any of the hypothesis testing techniques are always lower.

It is relevant at this point to mention that measuring regret in terms of ln $L_2$ instead of in terms of $L_2$ tends to give more advantage to more parsimonious model selection strategies. The fact that model 1.00, which requires no estimation, is a possible model creates an oddity that a strategy that always chooses that model will have minimax regret in terms of ln $L_2$ compared to other strategies. This arises since when that model is the true model, that strategy will have $L_2$=0, so the positive $L_2$ of other strategies will result in the maximum regret of using them being in essence infinitely worse when considering differences in ln $L_2$ (of course ln $L_2$ is not calculable then). Notably, having model 1.00 as the true model does *not* provide the worst-case ln $L_2$ scenario for hypothesis testing strategies against the other strategies (this is fortunate since it is not desirable to have this unusual case drive the results), although it does provide the worst-case ln $L_2$ scenario for AIC,

AICc, AICu, and CV against SIC when there is known to be no trend (as in Table 5) or when the trend status is unknown, as in Table 7 which will be introduced immediately.[30]

Tables 6 and 7 are analogous to Table 5 in that they are dealing with the maximum regret on the $\ln L_2$ distance. They differ only in that Table 6 deals with the situation in which the researcher has correct prior knowledge that there is a trend in both variables, and that Table 7 deals with the situation in which the researcher does not have prior knowledge about the trend status of the variables. As in Table 5, SIC is the best performer in minimizing the maximum regret in Tables 6 and 7.

**Table 6. Maximum regret on ln $L_2$, trend in *Y* and *Z* in data generating process, only models with a trend are choosable[a]**

| | AIC | AICc | AICu | SIC | CV | EG-10% | EG-5% | EG-10/5 | Jo-10% | Jo-5% | Jo-10/5 |
|---|---|---|---|---|---|---|---|---|---|---|---|
| AIC | 0.00 | 0.19 | 0.47 | 0.61 | 0.21 | 0.76 | 0.90 | 0.77 | 0.83 | 1.06 | 0.94 |
| AICc | 0.18 | 0.00 | 0.32 | 0.51 | 0.28 | 0.68 | 0.81 | 0.68 | 0.74 | 0.96 | 0.84 |
| AICu | 0.41 | 0.25 | 0.00 | 0.19 | 0.52 | 0.41 | 0.55 | 0.42 | 0.43 | 0.65 | 0.80 |
| SIC | 0.42 | 0.26 | 0.07 | 0.00 | 0.53 | 0.34 | 0.41 | 0.30 | 0.35 | 0.47 | 0.82 |
| CV | 0.21 | 0.39 | 0.66 | 0.70 | 0.00 | 0.85 | 0.95 | 0.82 | 0.89 | 1.11 | 0.97 |
| EG-10% | 13.41 | 13.41 | 13.41 | 13.41 | 13.41 | 0.00 | 6.08 | 6.07 | 11.24 | 11.39 | 11.30 |
| EG-5% | 12.66 | 12.66 | 12.66 | 12.66 | 12.66 | 0.86 | 0.00 | 0.15 | 10.53 | 10.68 | 10.59 |
| EG-10/5 | 12.77 | 12.9 | 13.08 | 13.11 | 12.74 | 0.86 | 6.08 | 0.00 | 11.23 | 11.38 | 11.29 |
| Jo-10% | 6.64 | 6.64 | 6.64 | 6.64 | 6.63 | 0.67 | 0.69 | 0.70 | 0.00 | 1.19 | 1.08 |
| Jo-5% | 5.81 | 5.81 | 5.82 | 5.82 | 5.81 | 0.86 | 0.60 | 0.61 | 0.86 | 0.00 | 1.30 |
| Jo-10/5 | 6.42 | 6.43 | 6.43 | 6.43 | 6.42 | 0.86 | 0.32 | 0.00 | 0.86 | 1.18 | 0.00 |

[a] Each cell's value shows the maximum regret (greatest increase in the ln $L_2$ distance) when using the procedure listed for the cell's row rather than using the procedure listed for the cell's column. The shaded cells are those associated with the strategy that has minimax regret compared to all other strategies.

**Table 7. Maximum regret on ln $L_2$, both models with a trend and models without a trend are used in data generating process, all models in Table 1 choosable[a]**

| | AIC | AICc | AICu | SIC | CV | EG-10% | EG-5% | EG-10/5 | Jo-10% | Jo-5% | Jo-10/5 |
|---|---|---|---|---|---|---|---|---|---|---|---|
| AIC | 0.00 | 0.22 | 0.65 | 1.03 | 0.20 | 0.87 | 1.23 | 1.10 | 0.87 | 1.23 | 1.10 |
| AICc | 0.17 | 0.00 | 0.52 | 0.89 | 0.27 | 0.77 | 1.10 | 0.97 | 0.77 | 1.10 | 0.97 |
| AICu | 0.41 | 0.25 | 0.00 | 0.37 | 0.48 | 0.51 | 0.72 | 0.54 | 0.51 | 0.72 | 0.54 |
| SIC | 0.44 | 0.30 | 0.12 | 0.00 | 0.53 | 0.51 | 0.55 | 0.43 | 0.52 | 0.56 | 0.45 |
| CV | 0.26 | 0.47 | 0.85 | 1.09 | 0.00 | 0.93 | 1.30 | 1.17 | 0.93 | 1.3 | 1.17 |
| EG-10% | 13.58 | 13.61 | 13.68 | 13.71 | 13.59 | 0.00 | 5.97 | 5.74 | 11.19 | 11.34 | 11.26 |
| EG-5% | 12.83 | 12.86 | 12.93 | 12.95 | 12.84 | 1.49 | 0.00 | 0.84 | 10.51 | 10.65 | 10.57 |
| EG-10/5 | 12.9 | 12.92 | 12.99 | 13.02 | 12.91 | 2.55 | 5.34 | 0.00 | 11.17 | 11.32 | 11.24 |
| Jo-10% | 7.72 | 7.77 | 7.85 | 7.86 | 7.70 | 0.30 | 0.36 | 0.76 | 0.00 | 1.24 | 2.39 |
| Jo-5% | 7.32 | 7.34 | 7.38 | 7.40 | 7.32 | 1.19 | 0.27 | 0.61 | 1.12 | 0.00 | 2.40 |
| Jo-10/5 | 7.50 | 7.56 | 7.63 | 7.64 | 7.49 | 1.95 | 1.56 | 0.00 | 1.87 | 1.62 | 0.00 |

[a] Each cell's value shows the maximum regret (greatest increase in the ln $L_2$ distance) when using the procedure listed for the cell's row rather than using the procedure listed for the cell's column. The shaded cells are those associated with the strategy that has minimax regret compared to all other strategies.

[30] It is a concern for this study that at a low sample size, the information criteria investigated here can have difficulty in distinguishing a nonstationary process from a stationary one in comparison to typical unit-root tests, especially when there is a trend in the data generating process.

## 9. Inference and application in practice

The extensive use of information criteria to select among time series models, as the previous section's findings support, should be an aid to the researcher in finding a credible empirical model. However, once one finds an empirical model that is most supported by the data, as indicated by the minimization of an information criterion, there still remains the question of whether the data strongly or weakly supports that model over the others. One could try to complement a model chosen through information criterion minimization with more standard hypothesis tests and confidence intervals to get at this issue of how strongly the data support a particular model. However, the legitimacy of the nominal sizes of hypothesis tests and the nominal degree of confidence for each confidence interval may be seriously questioned as the data-driven pre-selection of the model tested would likely affect actual sizes and actual degrees of confidence. Alternatively, one could calculate a weight for each model *i*, using the formula

$$w_i \equiv \frac{\exp(-\frac{1}{2}\Delta_i)}{\sum_{r=1}^{R}\exp(-\frac{1}{2}\Delta_r)}, \qquad (17)$$

where $\Delta_i \equiv IC_i - \min IC$, $IC_i$ represents the value of the information criterion for model *i*, min *IC* is the value of the information criterion for the choosable model that minimized the information criterion, and *R* is the number of models considered. In the case of AIC and AICc this weight arguably represents the degree of evidence supporting model *i* being the "K-L best" model (the model that minimizes that Kullback-Leibler (1951) distance) from among the *R* models.[31] In the case of SIC, this weight arguably represents the weight of evidence supporting model *i* being the model with the highest posterior probability of being the "quasi-true" model, i.e. the one which SIC would choose asymptotically, from among the *R* models (Burnham and Anderson, 2002). The weight $w_i$ can be calculated for each model and the models ranked according to the magnitude of the weight.

It is inappropriate to rely upon just information criteria and/or hypothesis testing in investigation for empirical studies, as a little extra thought and checking of residuals by the researcher can go a long way in avoiding erroneous conclusions. Hendry and Richard (1983) suggested six criteria that should be met for a chosen model to be acceptable. We repeat them here and discuss them in the context of using information criteria extensively in choosing among time series models as we have advocated.

[31] The Kullback-Leibler (1951) distance measures how close a probability function is to the true probability model which generates some data. Akaike (1978) advocated interpreting exp(-0.5×$IC_i$) multiplied by a constant as the likelihood of model *i* being the K-L best model, a position supported by Bozdogan (1987) among others. This interpretation implies that exp(-0.5$\Delta_i$) reflects the likelihood of model *i* being the K-L best model relative to that likelihood for the model that minimizes the information criterion (as noted in Akaike (1983)) and $w_i$ reflects the probability that out of all the models considered, model *i* is the K-L best model (Burnham and Anderson, 2002).

One criterion for a chosen model to be acceptable is that the chosen model should be *data-admissible*, allowing logical predictions, and another criterion is that the chosen model should be *theory-consistent*. The practitioner should for example check to make sure the signs on the coefficient estimates are logical and as a group are not leading to unusual predictions for the dependent variable. A third criterion is that the explanatory variables should *display weak exogeneity*. In the context of time series models this is an exceptionally important issue since in particular with macroeconomic data, there is considerable feedback among many of the variables. If one is investigating the relation between two time series variables and such exogeneity does not exist, then considering only reduced forms of vector autoregressive (VAR) models rather than the single equation models of this paper would be a suggested alternative, and since information criteria exist for choosing among VAR models, the methodology we have suggested here could conceivably work acceptably in extension to that environment.

A fourth criterion suggested by Hendry and Richard is that the *parameter estimates should display constancy*. The degree to which parameters estimates vary between the chosen model and the most closely competing models can provide information on how much confidence we should have in those estimates. We suggest that the higher the competing models are in their strength of evidence weights, $w_i$, the more attention we should pay to how their parameter estimates differ from those of the chosen model. Lack of parameter constancy can seriously diminish predictive reliability. One promising way of improving predictive reliability under such circumstances is through model averaging based on the strength of evidence weights, i.e. finding a new estimated model by averaging over each considered parameter across models, with the average being a weighted mean based on the strength of evidence weight, $w_i$ (Burnham and Anderson, 2002; Akaike, 1978, also had some early ideas along these lines).

A fifth criterion that Hendry and Richard found important is that the chosen model should *be data-coherent,* i.e. patterns in the residuals should not exist. Visual inspection of the residuals of the chosen model based is a powerful tool in this regard. One could also extend the use of information criteria to look for patterns in the residuals. The researcher could for example estimate the equation used for the Breusch-Godfrey test of first-order autocorrelation, estimate it again under the null hypothesis of no first-order autocorrelation, and use an information criterion to decide which of the two estimated equation is more supported by the data.

The last criterion of Hendry and Richard is that the chosen model should be *encompassing*, i.e. be able to explain the results of rival models. The chosen model using an information criteria tends to be superior at explaining the data at hand than the other considered models in the sense of approximately better predictive performance with that data. This is particularly the case with AIC (or AICc or AICu) as they are built to find the model that is K-L best. The simulations in this paper suggest that SIC leads to good predictive results also in terms of ln $L_2$, although predictive capability is not its explicit aim. The issue of encompassing is likely more problematic with hypothesis testing, since comparisons between non-nested models is more difficult then.

## 10. Conclusions

Since time series data analysis is used enormously in empirical studies more research on the important issue of model selection is warranted. It is important to take into account model uncertainty since there might be many potential models. This issue seems to be frequently neglected by practitioners currently in empirical studies. It is common practice to present and rely on a single model that the practitioner has tediously ended up with by multiple steps of hypothesis testing. Since many of the models considered are usually nested it means that the problem of mass significance might very well exist.

In this paper we suggest using minimization of an information criterion more extensively for model selection in a time series environment. Our simulations show that this procedure often works well and better than hypothesis testing approach in choosing an appropriate model. Given the goal of an information criterion compared to a hypothesis test, this is perhaps not surprising, but it is also not so obvious given the complexities of issues involved with time series data, particularly those dealing with possible nonstationarity of the data. The use of an information criterion also has additional advantages in that it is simple to use and it can rank potential models based on how much the data support each model. How much the data support a particular model can be estimated through a weight calculated by considering the difference between the magnitude of the information criterion for a model and that magnitude for the model in which that information criterion is minimized. This weight may be calculated for each model considered and the resulting weights can be used to average the parameters across models, resulting perhaps in a new model with predictive reliability that is superior to that of the single model most supported by the data.

**Appendix: Converting Single Equation Processes into Associated VAR Processes**

In this paper various scenarios will be simulated, limited to the three broad categorizations of two variables with no cointegration, two variables with cointegration, and three variables with cointegration.

In the two-variable case when there is no cointegration, the matrix representation is given by

$$\begin{bmatrix}\Delta Y_t \\ \Delta Z_t\end{bmatrix} = \begin{bmatrix} b_1 & b_2 \\ m_1 & m_2 \end{bmatrix}\begin{bmatrix}1 \\ t\end{bmatrix} + \begin{bmatrix} 0 & b_6 \\ 0 & 0 \end{bmatrix}\begin{bmatrix}Y_t \\ Z_t\end{bmatrix} + \begin{bmatrix} b_3 & b_7 \\ m_4 & m_3 \end{bmatrix}\begin{bmatrix}Y_{t-1} \\ Z_{t-1}\end{bmatrix}$$
$$+ \begin{bmatrix} b_4 & b_9 \\ 0 & 0 \end{bmatrix}\begin{bmatrix}\Delta Y_{t-1} \\ \Delta Z_{t-1}\end{bmatrix} + \begin{bmatrix} b_5 & b_{10} \\ 0 & 0 \end{bmatrix}\begin{bmatrix}\Delta Y_{t-2} \\ \Delta Z_{t-2}\end{bmatrix} + \begin{bmatrix} w_{Yt} \\ w_{Zt}\end{bmatrix}$$

or equivalently,

$$X_t = N_D D_t + N_0 X_t + N_1 X_{t-1} + N_2 X_{t-2} + N_3 X_{t-3} + w_t, \qquad \text{(A.1)}$$

where

$$X_t = \begin{bmatrix}Y_t \\ Z_t\end{bmatrix},\ D_t = \begin{bmatrix}1 \\ t\end{bmatrix},\ N_D = \begin{bmatrix} b_1 & b_2 \\ m_1 & m_2 \end{bmatrix},\ N_0 = \begin{bmatrix} 0 & b_6 \\ 0 & 0 \end{bmatrix},\ N_1 = \begin{bmatrix} 1 + b_3 + b_4 & b_7 + b_9 \\ m_4 & m_3 + 1 \end{bmatrix},$$

$$N_2 = \begin{bmatrix} b_5 - b_4 & b_{10} - b_9 \\ -m_4 & 0 \end{bmatrix}, \text{ and } N_3 = \begin{bmatrix} -b_5 & -b_{10} \\ 0 & 0 \end{bmatrix}.$$

Solving (A.1) for $X_t$, we get

$$X_t = \psi D_t + M_1 X_{t-1} + M_2 X_{t-2} + M_3 X_{t-3} + (I - N_0)^{-1} w_t \qquad \text{(A.2)}$$

where $\psi = (I - N_0)^{-1} N_D$ , $M_i = (I - N_0)^{-1} N_i$ for $i$ = 1, 2, 3.

Equation (A.2) is used to generate the data in this situation.

In the two-variable case where there is cointegration, the matrix representation is given by

$$\Delta X_t = \Phi D_t + \alpha[\rho', \beta'] \begin{bmatrix} D_t \\ \cdots \\ X_{t-1} \end{bmatrix} + \Gamma_1 \Delta X_{t-1} + \Gamma_2 \Delta X_{t-2} + w_t, \tag{A.3}$$

where

$\Delta X_t = X_t - X_{t-1}$, $\Phi = \begin{bmatrix} b_1 & b_3 \\ m_1 & m_2 \end{bmatrix}$, $\alpha = \begin{bmatrix} b_{11} & 0 \\ 0 & 1 \end{bmatrix}$, $\rho = \begin{bmatrix} -c_1 & 0 \\ 0 & 0 \end{bmatrix}$, $\beta = \begin{bmatrix} 1 & 0 \\ -c_2 & m_3 \end{bmatrix}$, $\Gamma_1 = \begin{bmatrix} b_4 & b_9 \\ m_4 & 0 \end{bmatrix}$

, and $\Gamma_2 = \begin{bmatrix} b_5 & b_{10} \\ 0 & 0 \end{bmatrix}$. The first columns in $\rho$ and $\beta$ constitute together (i.e. the first row of $[\rho', \beta']$ ) the only possible cointegrating vector; the second columns on those matrices do not provide a cointegrating vector since there is at most only one non-zero parameter in it.

Solving (A.3) for $X_t$ we get

$$X_t = \psi D_t + M_1 X_{t-1} + M_2 X_{t-2} + M_3 X_{t-3} + w_t, \tag{A.4}$$

where $\psi = \Phi + \alpha\rho'$, $M_1 = I + \alpha\beta' + \Gamma_1$, $M_2 = \Gamma_2 - \Gamma_1$, and $M_3 = -\Gamma_2$

Equation (A.4) is used to generate the data in this situation.